\documentclass[ba,preprint]{imsart}
\volume{TBA}
\issue{TBA}
\arxiv{2203.11664}
\firstpage{1}
\lastpage{28}

\usepackage{amsthm}
\usepackage{amsmath}
\usepackage{natbib}

\usepackage{xr-hyper}  

\usepackage[colorlinks,citecolor=blue,urlcolor=blue,filecolor=blue,backref=page]{hyperref}
\usepackage{graphicx}

\startlocaldefs
\usepackage{amssymb,amsfonts,dsfont,microtype,tikz}
\usetikzlibrary{calc,decorations.pathreplacing,calligraphy,fit}

\newcommand{\add}[1]{{#1}}  

\endlocaldefs

\begin{document}


\begin{frontmatter}
\title{Bayesian Learning of Graph Substructures\support{This work was supported by the Singapore Ministry of Education Academic Research Fund Tier~2 under Grant MOE2019-T2-2-100.}}
\runtitle{Bayesian Learning of Graph Substructures}

\begin{aug}
\author{\fnms{Willem} \snm{van den Boom}\thanksref{NUS}\ead[label=eW]{vandenboom@nus.edu.sg}},
\author{\fnms{Maria} \snm{De Iorio}\thanksref{NUS,ASTAR,UCL}\ead[label=eM]{mdi@nus.edu.sg}}
\and
\author{\fnms{Alexandros} \snm{Beskos}\thanksref{UCL,ATI}\ead[label=eA]{a.beskos@ucl.ac.uk}}

\runauthor{W.\ van den Boom et al.}

\address[NUS]{
Yong Loo Lin School of Medicine, National University of Singapore
\printead{eW}
\printead*{eM}
}

\address[ASTAR]{
Singapore Institute for Clinical Sciences, Agency for Science, Technology and Research
}

\address[UCL]{
Department of Statistical Science, University College London
\printead{eA}
}

\address[ATI]{
Alan Turing Institute, UK
}


\end{aug}

\begin{abstract}  
Graphical models provide a powerful methodology for learning the conditional independence structure in multivariate data. Inference is often focused on estimating individual edges in the latent graph. Nonetheless, there is increasing interest in inferring more complex structures, such as communities, for multiple reasons, including more effective information retrieval and better interpretability. Stochastic blockmodels offer a powerful tool to detect such structure in a network. We thus propose to exploit advances in random graph theory and embed them within the graphical models framework. A consequence of this approach is the propagation of the uncertainty in graph estimation to large-scale structure learning. We consider Bayesian nonparametric stochastic blockmodels as priors on the graph. We extend such models to consider clique-based blocks and to multiple graph settings introducing a novel prior process based on a \add{D}ependent Dirichlet process. Moreover, we devise a tailored computation strategy of Bayes factors for block structure based on the Savage-Dickey ratio to test for presence of larger structure in a graph. We demonstrate our approach in simulations as well as on real data applications in finance and transcriptomics.
\end{abstract}


\end{frontmatter}

\section{Introduction}
\label{sec:intro}

Graphical models provide a flexible tool to describe the conditional independence structure in multivariate data:
the nodes of the graph represent variables and the edges amongst them define conditional dependence \citep{Lauritzen1996}.
Most inferential approaches focus on estimation of individual edges rather than on identification of informative structure in a graph on a larger scale.
This is despite the fact that such large-scale structure is often present and of interest in multivariate data \citep{Ravasz2002,Yook2004}. Moreover, estimation of a single edge is (often extremely) sensitive to the number of observations as well as to the presence of specific nodes in the graph.
We therefore propose graphical models that also enable learning of large-scale structure.
These models build on extensive work in random graphs and networks \citep{Newman2011,Fortunato2016,Lee2019}
such as the stochastic blockmodel \citep{Holland1983}.

It is important to stress the distinction between random graph theory and graphical models as two parallel, large research areas with only limited interplay such as the work by \citet{Bornn2011} and \citet{Peixoto2019}.
Within the first field, models for random graphs have evolved substantially from initial approaches such as the Erdős–Rényi model \citep{Erdos1959} to methods for large-scale structure.
They usually involve the description of network formation/evolution. See
\citet{Fienberg2012} \add{and \citet{Barabasi2016}} for an overview.
Such developments contrast with the literature on graphical models \add{\citep{Friedman2007,Armstrong2009,
Zhou2011,Maathuis2019,Ni2022}} that aims to infer a graph from multivariate data. In this context, focus of inference is usually to determine the presence of an edge between two nodes whereas modelling of large-scale graph structures is often neglected \citep{Bornn2011}.

The rationale underpinning our work derives from the following consideration. In the random graph literature,
there is major interest on large-scale structures as they often arise in applications. A common example
is provided by
scale-free networks which imply a hub structure \citep{Yook2004}. More recently, \citet{Newman2011} advocates for more complex structures such as modularity.
This consideration motivates the need to investigate such components also when inferring graphs from multivariate data. On the other hand, the Bayesian graphical model literature
commonly focuses on single edges, and specification of a prior on graph space is achieved assuming the same probability of inclusion for each edge with all edges being independent. This prior
corresponds to the Erdős–Rényi model.

Focusing on single edges can be restrictive in many applications, often preventing detection of important data features.
For instance, assume we are interested in estimating a graph from gene expression data. It could be of biological interest (e.g.\ disease aetiology) to group genes in co-expression modules (i.e.~block, larger structure) \citep{Yook2004}. Similarly in metabolomics, it is of interest to identify metabolites that are involved in the same biochemical reaction/pathway \citep{Ravasz2002}.
Social networks provide another area where such graph substructures are relevant. For instance, we might estimate a graph from voting records of members of parliament with the goal of identifying political factions.

The increasing interest in estimating large structures in multivariate data is reflected in recent work. For example, \cite{Zhang2018} first estimates a graph from the data and, then, identifies large-scale structure using random graph
methods. Such an approach is suboptimal, for instance because it does not propagate the uncertainty from
network estimation to the estimation of large-scale structure. In the machine learning literature,
methods for identification of graph substructure can be found in \citet{Marlin2009b}.

Our work is positioned in this new line of research. Exploiting advances from random graph theory, we propose graphical models able to
accommodate single-edge as well as block structure.
The benefits of joint graph and structure recovery compared to a two-step approach are multiple:
(i) if present, large-scale structure can guide
graph estimation; (ii) ad hoc specification of a graph estimator (e.g.\ through an edge inclusion
probability threshold) is not required; (iii) data-driven detection of structure or lack
thereof; (iv) uncertainty in graph estimation
propagates to large-scale structure learning; (v) extension to complex set-ups (e.g.\ different biological conditions) is in principle straightforward which leads to; (vi) effective use of information as the developed framework allows combining data from multiple
sources in a principled way.
We consider both single and multiple graph scenarios
as well as different blockmodels,
namely the usual stochastic blockmodel and also one where
blocks are cliques.
Here we focus on blockmodels because of their popularity, but the setting is general and other priors could be employed.

One of our contributions is an algorithm (derived as by-product of the MCMC) to
compute Bayes factors to test for the presence of block structure, which is equivalent to
the presence of clusters in a nonparametric partition model. Our approach, based on the Savage-Dickey ratio \citep{Dickey1971}, offers computational advantages over existing methods \citep{Basu2003,Legramanti2022_1}.
To define blocks in multiple graphs, we introduce a novel Bayesian nonparametric prior. Specifically, we propose a \add{D}ependent Dirichlet process
that does not enforce exchangeability within groups as in previous approaches
\add{\citep[e.g.][]{MacEachern1999,DeIorio2004,Muller2004,Camerlenghi2019,Quintana2022}}.

The paper is structured as follows.
Sections~\ref{sec:sbm_review} and \ref{sec:ggm_review}
review related work on blockmodels and graphical models, respectively.
Section~\ref{sec:ggm} introduces Gaussian graphical models \citep[GGMs,][]{Dempster1972}. In
Section~\ref{sec:graph_prior}, we propose various priors on graphs that allow recovery of large-scale structure.
Section~\ref{sec:test} introduces Bayes factors for testing for block structure. We demonstrate the proposed approach in simulation studies in 
Section~\ref{sec:simul} and on real data applications in Section~\ref{sec:applications}.
We conclude the paper in Section~\ref{sec:discussion}.
\add{The paper is accompanied by Supplementary Material which contains further details on the methods and results.}

\subsection{Stochastic Blockmodels}
\label{sec:sbm_review}

Arguably, the most widely used model for large-scale structure in graphs is the blockmodel
\citep{Fienberg2012}
which is therefore our starting point.
A stochastic blockmodel \citep{Holland1983} consists of a partition of the set of nodes into blocks or communities, where we use both terms interchangeably.
Then, nodes in the same block are more likely to be connected than
nodes from different blocks.
Thus, the structure of interest is the clustering of nodes and the connectivity within and between these clusters.
Introducing block structure in
graph estimation allows highlighting macro-organisation (instead of focusing on single edges) and important hubs/connectivity clusters, which ultimately will aid interpretation of the results and hypothesis generation.

To this end, the key modelling strategy that we adopt
is to employ tools from the Bayesian nonparametric literature for estimation of block structure in a graph, as such a strategy provides uncertainty propagation across the full graphical model
and data-driven determination of blocks' number and membership.
\add{We limit this literature review of blockmodels to Bayesian nonparametric approaches though many others exist \citep[e.g.][]{Fortunato2016,Abbe2018,Funke2019,Lee2019,Gao2021}.}
See \citet{Schmidt2013} for an introduction to Bayesian nonparametric modelling of graphs including blockmodels.
\citet{Kemp2006} introduce a blockmodel where the prior on the partition of nodes is a
Chinese restaurant process
\citep[CRP,][]{Pitman2006} which closely relates to the Dirichlet process \citep[DP,][]{Ferguson1973}.
\citet{Geng2018}\add{, \citet{Gao2020} and \citet{Jiang2021}} place a mixture \add{prior with random number of components}
\citep{Miller2017,Argiento2022} on the partition\add{: \citet{Geng2018}}
obtain posterior consistency results for the number of blocks\add{
, and \citet{Jiang2021} do so for the partition and the edge probabilities.
\citet{Gao2020} provide posterior concentration rates for the edge probabilities and show that their posterior mean achieves the minimax rate.
}
\citet{Legramanti2022} employ Gibbs-type partition priors which generalise both the
CRP and the mixture \add{with random number of components}.

In general, these approaches require also specification of prior edge inclusion probabilities jointly with the block structure prior. For instance,
\citet{Kemp2006}, \citet{Geng2018} and \citet{Legramanti2022} place Beta distributions on the edge
probabilities,
\add{and \citet{Reyes2016} and \citet{Jiang2021} add structure by using different priors for within- and between-block edge probabilities,}
while \citet{Tan2019} use a DP to build a joint prior
on the partition of nodes and edge probabilities.
Additionally, they extend the model to
a degree-corrected blockmodel, i.e.~they introduce a popularity parameter for each node.
\citet{Passino2020} \add{and \citet{Loyal2022} consider Bayesian blockmodels} where edge probabilities derive from a latent space embedding.
\add{
\citet{Caron2017} model edge probabilities by
associating edges with realisations from a Poisson process with rate described by a random measure.
\citet{Herlau2016}
and
\citet{Todeschini2020}
extend this approach, respectively, to (i) blockmodels,
with blocks corresponding to subsets of the support of a random measure;
and (ii) to overlapping community detection, with each community corresponding to an element of a compound random measure.
}
The Bayesian nonparametric blockmodel by \citet{Peixoto2017} defines a generative process based on a random partition prior on the block configuration where structural constraints are imposed on the number of edges across blocks. This approach avoids explicit modelling of edge inclusion probabilities.

\subsection{Learning Block Structure in Graphical Models}
\label{sec:ggm_review}

Proposals for the estimation of large-scale structures in graphical models can be categorised in two main strategies: (i) regularisation methods; (ii) imposing structure on the precision matrix.
Within the first framework,
\citet{Ambroise2009} and \citet{Marlin2009b} do not model graphs explicitly, but learn a block structure as part of a shrinkage estimator for the precision matrix as in the graphical lasso \citep{Friedman2007}, where every block is characterised by its own regularisation parameter.

Within the second framework,
\citet{Sun2014} consider a GGM with a
CRP as prior on the partition of nodes.
Then, the partition informs the sparsity pattern of the scale matrix of the Wishart prior on the precision matrix rather than of the precision matrix itself as is commonly done in GGMs.
See Section~\ref{sec:sun} of Supplementary Material \add{\citep{vandenBoom2022_supp}} for details.
\citet{Marlin2009a} impose sparsity in the precision matrix of a GGM by first approximating the joint distribution of the nodes via the specification of the conditional distribution of each node given the others.
Then, they impose
a continuous spike-and-slab prior on ``edge weights'' that capture the association of a node with the others. Finally, the prior on edge weights incorporates a block structure.
\citet{Sun2015} fix the number of blocks, place a Dirichlet prior on the partition of the nodes in an exponential graphical model and compute a point estimate of the partition using an expectation-maximisation algorithm.

\citet{Bornn2011} consider decomposable graphs, which allow modelling of cliques and separators separately, and use a
product partition model as prior on the graph. Their prior can induce large cliques and, as such, allows the identification of larger structures than
edges.
\citet{Peixoto2019}
uses a stochastic blockmodel as prior for network reconstruction in two discrete-valued graphical models, i.e.~the Ising model and an epidemic model of infection status across time.
He shows empirically that joint estimation of the graph and block structure increases accuracy as compared to two-step approaches.
\add{\citet{Colombi2022} and \citet{Cremaschi2022} consider inference in GGMs under a known block structure with either all or no edges present between a pair of blocks.}

\section{Gaussian Graphical Models}
\label{sec:ggm}

Let the graph $G=(V,E)$ be defined by a set of edges $E \subset {\{(i,j)\mid 1\leq i<j\leq p \}}$
that represent links among the nodes in $V=\{1,\ldots,p\}$.
The data are represented by an
$n\times p$ matrix $Y$ with independent and identically distributed rows corresponding to $p$-dimensional random vectors whose elements are represented by nodes on the graph.
A graphical model \citep{Lauritzen1996} is a family of distributions on the rows which is Markov over $G$.
That is, the distribution $p(Y\mid G)$ is such that
the $i$-th and $j$-th columns of $Y$ are independent conditionally on the other columns if and only if
$(i,j)\notin E$.

While our development for learning
large-scale structure applies to graphical models in general,
here we focus on GGMs \citep{Dempster1972}, which consider a Gaussian law for $p(Y\mid G)$.
Then,
each row of $Y$ is distributed according to a Multivariate Gaussian distribution $\mathcal{N}(0_{p\times 1},\, \Omega^{-1})$
with precision matrix $\Omega$.
The conditional independence
structure implied by $G$ implies that
$\Omega_{ij} = 0$
if and only if nodes $i$ and $j$ are not connected.
For the complete matrix $Y$,
$\Omega_{ij} = 0$
implies that the $i$-th and $j$-th columns of $Y$ are independent
conditionally on the others.
In this context,
a blockmodel on $G$ enables learning of
sparse block-structured precision matrices where the block structure is unknown.

A popular choice as prior $p(\Omega\mid G)$ for the precision matrix $\Omega$ conditional on the graph $G$ is the $G$-Wishart distribution $\mathcal{W}_G(\delta, D)$ as it induces conjugacy and allows working with non-decomposable graphs \add{\citep{Giudici1996,Roverato2002}}. It is parameterised by degrees of freedom $\delta > 2$
and a positive-definite rate matrix $D$.
Then
\citep[e.g.][]{AtayKayis2005},
\begin{equation} \label{eq:ggm_lik}  
	p(Y\mid G)\propto \int p(\Omega\mid G)\, p(Y\mid \Omega)\, d\Omega
	= \frac{I_G(\delta^\star, D^\star)}{(2\pi)^{np/2} I_G(\delta, D)},
\end{equation}
where
$\delta^\star = \delta + n$,
$D^\star = D + Y^\top Y$ and
$I_G(\delta, D)$ is the normalising constant of the density of $\mathcal{W}_G(\delta, D)$.
The constant
$I_G(\delta, D)$ is not analytically available for general, non-decomposable $G$.
Thus, we make use of the Markov chain Monte Carlo (MCMC) methodology from \citet{vandenBoom2022}
and of a Laplace approximation of $I_G(\delta, D)$ from \citet{Moghaddam2009}
to perform posterior inference on $G$.

\section{Graph Priors for Large-Scale Structure Recovery}
\label{sec:graph_prior}

Key to learning large structure in graphs is specification of a prior $p(G)$ on graphs.
To this end, we borrow ideas from random graph theory, adapting them effectively in our context.

\add{Moreover, our approach is based on the
Dirichlet process \citep{Ferguson1973},
a probability model for random
probability distributions. Readers familiar with the DP can skip to the next subsection.
If a random measure $H\sim \mathrm{DP}(\nu,H_0)$, then $H$ is almost surely discrete. $H_0$ is the base measure, a distribution around which the DP is centred, while $\nu>0$ denotes the precision parameter. 
Due to its discreteness, $H$ admits the well-known ``stick-breaking'' construction \citep{Sethuraman1994} and can be represented as a countable mixture of point masses:
$H = \sum_{k=1}^{\infty} w_k \delta_{\beta'_k}$. Here $\delta_{\beta'_k}$ is a point mass at $\beta'_k$, the weights $w_k$ are generated by rescaled Beta distributions,
$w_k = \xi_k \prod_{l=1}^{k-1} (1-\xi_l)$
with $\xi_k \stackrel{\text{i.i.d.}}{\sim} \mathrm{Beta}(1,\nu)$,
and the locations $\{\beta'_k\}_{k=1}^\infty$ are i.i.d.\ samples from the base measure $H_0$.
Finally, the sequences $\{\beta'_k\}_{k=1}^\infty$ and $\{\xi_k\}_{k=1}^\infty$ are independent.}

\subsection{Degree-Corrected Stochastic Blockmodel}
\label{sec:sbm}

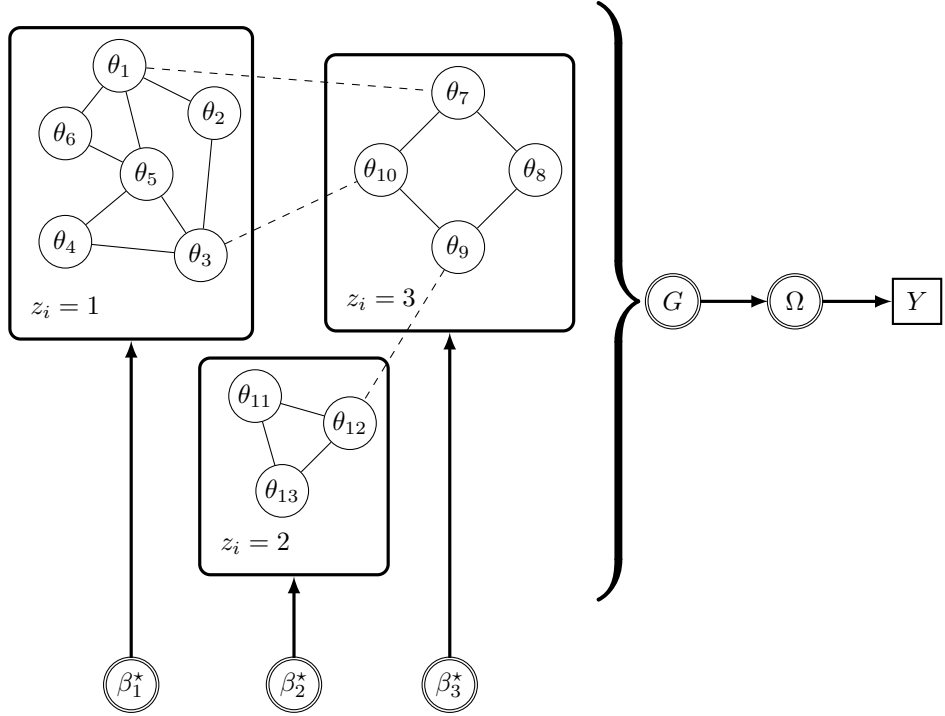
\begin{figure}[tb]
	\centering
	\begin{tikzpicture}[scale=1.8,
        node/.style={
            draw=black, minimum width=.7cm, inner sep=0, circle,
        },
        par_node/.style={node, double},
        box/.style={
            draw=black, very thick, rounded corners, inner sep=1.5mm
        },
        arrow/.style={-latex, very thick}
    ]
        
		\node[node] (1) {$\theta_1$};
        \node[node] (2) at ([shift={(.7, -.35)}]1) {$\theta_2$};
        \node[node] (5) at ([shift={(.2, -.8)}]1) {$\theta_5$};
        \node[node] (6) at ([shift={(-.4, -.5)}]1) {$\theta_6$};
        \node[node] (3) at ([shift={(.4, -.6)}]5) {$\theta_3$};
        \node[node] (4) at ([shift={(-.6, -.5)}]5) {$\theta_4$};

        \newdimen\offset
        \offset=.57cm
        \node[node] (7) at ([shift={(2.5, -.2)}]1) {$\theta_7$};
        \node[node] (8) at ([shift={(\offset, -\offset)}]7) {$\theta_8$};
        \node[node] (10) at ([shift={(-\offset, -\offset)}]7) {$\theta_{10}$};
        \node[node] (9) at ([shift={(\offset, -\offset)}]10) {$\theta_9$};

        \node[node] (11) at ([shift={(-1.5, -1.1)}]9) {$\theta_{11}$};
        \node[node] (12) at ([shift={(.7, -.2)}]11) {$\theta_{12}$};
        \node[node] (13) at ([shift={(.2, -.7)}]11) {$\theta_{13}$};
        
        \draw (1) -- (2);
		\draw (1) -- (6);
        \draw (1) -- (5);
        \draw (2) -- (3);
        \draw (3) -- (4);
        \draw (3) -- (5);
        \draw (4) -- (5);
        \draw (5) -- (6);

        \draw (7) -- (8);
        \draw (8) -- (9);
        \draw (9) -- (10);
        \draw (10) -- (7);

        \draw (11) -- (12);
        \draw (12) -- (13);
        \draw (13) -- (11);

        \draw[dashed] (1) -- (7);
        \draw[dashed] (3) -- (10);
        \draw[dashed] (9) -- (12);

        \offset=-.19cm
        \node at ($(4.center |- 3.south) + (0, \offset)$) (z1_label) {$z_i=1$};
        \node at ($(10.center |- 9.south) + (0, \offset)$) (z2_label) {$z_i=3$};
        \node at ($(11.center |- 13.south) + (0, \offset)$) (z3_label) {$z_i=2$};
        \node[box,fit=(1) (2) (3) (6) (z1_label)] (z1) {};
        \node[box,fit=(7) (8) (9) (10) (z2_label)] (z2) {};
        \node[box,fit=(11) (12) (z3_label)] (z3) {};

        \node[par_node] at ([shift={(0, -.8)}]z3.south) (beta3) {$\beta_2^\star$};
        \node[par_node] at (z1.south |- beta3) (beta1) {$\beta_1^\star$};
        \node[par_node] at (z2.south |- beta3) (beta2) {$\beta_3^\star$};
        \node[fit=(z1) (z2) (z3), inner sep=.3cm] (graph) {};
        \node[par_node] at ([shift={(.55, 0)}]graph.east) (G) {$G$};
        \node[par_node] at ([shift={(.9, 0)}]G) (Omega) {$\Omega$};
        \node[draw=black, inner sep=.18cm, thick] at ([shift={(.9, 0)}]Omega) (Y) {$Y$};

        \draw[arrow] (beta1) -- (z1.south);
        \draw[arrow] (beta2) -- (z2.south);
        \draw[arrow] (beta3) -- (z3.south);
        \draw[arrow] (G) -- (Omega);
        \draw[arrow] (Omega) -- (Y);

        \draw[decorate, decoration={calligraphic brace,amplitude=15pt}, line width=3pt] (graph.north east) -- (graph.south east);
        
	\end{tikzpicture}
	\caption{\add{Visualisation of the degree-corrected stochastic blockmodel. Each node is characterised by its own popularity parameter $\theta_i$. 
	Block membership is captured by the allocation parameter $z_i$.
    Nodes in the same block share the same value for $z_i$ as well as for the  interaction parameter $\beta_i=\beta_{z_i}^\star$. There are more edges within (solid lines) than between (dashed lines) blocks represented by square boxes. The resulting graph $G$ constrains the precision matrix $\Omega$ of the data $Y$.
    The direction of the arrows reflects the hierarchical specification of the model.}} \label{fig:sbm}
\end{figure}

\add{The fundamental idea behind our strategy is the following. Each node $i$ in the graph forms a connection with another node $j$ according to (i) its own propensity (or popularity) captured by the parameter $\theta_i$; (ii) its block membership captured by the``interaction'' parameter $\beta_{ij}$, with nodes in the same block having higher probability to share an edge. The popularity parameter can be thought of as the node-specific propensity to form connections with other nodes. To guarantee parsimony, popularity parameters are modelled assuming a DP prior. On the other hand, prior specification on  block-specific parameters is more complex, as it depends on the  prior on the partition of nodes, which defines the block structure. We exploit the discreteness of the DP to define the blocks, where each component in the  DP discrete mixture  (see the ``stick-breaking construction'') corresponds to a block and nodes allocated to the same component share the same block-specific interaction parameter. 
Figure~\ref{fig:sbm} summarises the modelling strategy.}

Our starting point is the Bayesian nonparametric degree-corrected stochastic blockmodel by \citet{Tan2019}
who propose a probit model for the edge inclusion probabilities. More specifically,
$\mathrm{Pr}\{(i,j)\in E\} = \Phi(\mu_{ij})$, independently over distinct pairs $(i,j)$ for $1\leq i<j\leq p$, where $\Phi(\cdot)$ is the \add{cumulative distribution function} of the standard Gaussian distribution $\mathcal{N}(0,1)$.
Then,
\begin{equation}
\mu_{ij} = \theta_i + \theta_j + \beta_{ij}\,
\mathds{1}
[z_i=z_j]
\label{eq:mu}
\end{equation}
The allocation variable $z_i$ denotes the community node $i$ belongs to.
The parameter $\beta_{ij}$ measures the strength of interaction among members of the same community with nodes in the same block expected to share more edges among themselves than with nodes outside. 
The popularity parameter $\theta_i$
allows for degree correction in the blockmodel.
That is, nodes have varying popularity as captured by the number of their neighbours, i.e.~the number of nodes they are connected to via an edge.

\add{To specify a prior on $\beta_{ij}$,} we introduce auxiliary variable $\beta_i$ for each node $i$ and assume $\beta_{i}\mid H \overset{\text{i.i.d.}}{\sim} H$ for $i\in V$ where $H \sim \mathrm{DP}(\nu, H_0)$
with $H_0 = \mathcal{N}(0, s^2_\beta)$, $s^2_\beta>0$. The discreteness of the DP implies a positive probability of ties in a sample from \add{$H = \sum_{k=1}^{\infty} w_k \delta_{\beta'_k}$} and this, in turn, induces a clustering structure so that the nodes will be grouped together in $K$ clusters, where the number $K$ of clusters is unknown and learn\add{t} from the data through the posterior distribution. In our context, each cluster corresponds to a block, the parameters $w_k$ denote the prior probabilities of belonging to each mixture component, and $\beta'_k$ denotes the block-specific interaction parameter. 
Nodes are clustered based on their edge inclusion probability, so that nodes in the same block $k$ share a common value $\beta'_k$ such that in~\eqref{eq:mu} $ \beta_{ij}=\beta_i=\beta_j=\beta'_k$.
Thus, the set of node-level parameters $\{\beta_{i}\}_{i=1}^p$ reduces a posteriori to the set of unique values $\beta_1^\star, \ldots, \beta_K^\star$ assigned to within-block edges.
We denote block membership 
with the variable $z_i \in \{1,\ldots K\}$, $i \in V$.

\add{
Each node is characterised by its own popularity parameter $\theta_i$, which allows for higher flexibility, but can also challenge identifiability of large-scale structures. This is due to the fact that some connection patterns might be explained nearly as well (in terms of graph likelihood) by either granular specification of the vector $\theta$ or by blocks.
To avoid such identifiability issues and associated inflated posterior uncertainty,
we reduce the number of unique elements of $\theta$
through clustering via a DP.
Specifically,}
$\theta_i\mid F \stackrel{\text{i.i.d.}}{\sim} F$ for $i\in V$,
where $F\sim\mathrm{DP}(\alpha, F_0)$, $\alpha>0$,
with $F_0 = \mathcal{N}(0, s^2_\theta)$, $s^2_\theta>0$.
Lastly,
$\nu\sim \Gamma(a_\nu,\, b_\nu)$
and
$\alpha\sim \Gamma(a_\alpha,\, b_\alpha)$, $a_\nu, b_\nu, a_\alpha, b_\alpha>0$.

\add{The proposed} model has wide applicability as blocks can represent communities.
Algorithm~\ref{alg:sbm} in Supplementary Material \add{\citep{vandenBoom2022_supp}} details an MCMC algorithm for posterior inference.
We remark that the block-specific parameter $\beta^\star_k$ does not appear
in the likelihood for a block
$k$ of size one, i.e.~when $|\{i\mid z_i = k\}| = 1$,
differently from conventional applications of the DP.
Therefore,
the Metropolis-Hastings proposal to add a block of size one does not require a proposal for the parameter specific to the new block.
This property
causes
Algorithms~2 and 8 in \citet{Neal2000}, which we use to update the DP parameters, to coincide.

\subsection{Southern Italian Community Structure}
\label{sec:sics}

In the stochastic blockmodel,
nodes in the same block are not necessarily connected.
This level of flexibility is particularly desirable when the network is observed directly, and focus is on understanding network formation and evolution.
On the other hand, in graphical models, the graph is a latent variable and, in applications, it might be appropriate to impose a more restrictive definition of block/community.
More restrictions can also provide the benefit of improving computations as the space to explore gets smaller.

Here we assume that nodes in a block form a clique.
In a clique, all pairs of nodes are connected by an edge.
Indeed, the earliest approaches
for community structure in graphs consider cliques \citep{Luce1949,Festinger1949},
the rationale being that
a community is strongest if
each pair of its members is connected.
Cliques have for instance biological significance in protein-protein interaction networks \citep{Yu2006}.
We thus introduce a Bayesian nonparametric graph prior where each block is
a clique.

In this context, main object of inference are
the cluster allocation variables $z_i$, $i\in V$,
as in Section~\ref{sec:sbm}, as then the block structure is given: $\mathrm{Pr}\{(i,j)\in E\} = 1$ if $z_i = z_j$.
We name such construction the Southern Italian community structure (SICS) 
with reference to traditional Southern Italian communities where everybody knows each other.
Still, there can be connections between nodes from different blocks and, for these edges, we assume a prior inclusion probability $\rho$:
$\mathrm{Pr}\{(i,j)\in E\} = \rho$
for $1\leq i<j\leq p$ and
$z_i \ne z_j$.
This construction defines a prior $p(G\mid z)$.

A prior on $z$ completes the graph prior $p(G)= \int p(G\mid z)\, p(z) dz$.
We assume that $z$ follows the Chinese restaurant process \citep{Pitman2006}
with concentration parameter $\nu$
a priori
for concreteness and consistency with the DP in Section~\ref{sec:sbm}. We note that our approach is flexible and other priors on the partition of nodes
$z=(z_1,\ldots,z_p)$
can be straightforwardly adopted borrowing from the rich Bayesian nonparametric literature.
The CRP assumption implies:
\begin{equation} \label{eq:dir}
	\mathrm{Pr}(z_i=k\mid z_{-i}) =
	\begin{cases}
		\frac{n^\beta_{-i,k}}{p - 1 + \nu}, &k=1,\dots,K^{-i} \\
		\frac{\nu}{p - 1 + \nu}, &k=K^{-i}+1
	\end{cases}
\end{equation}
where $z_{-i}=\{z_j\mid j\ne i\}$, $n^\beta_{-i,k} = |\{j\in V \mid z_j = k,\, j\ne i\}|$
and $K^{-i}$ is the number of unique elements in $z_{-i}$.
Finally,
$\rho\sim\mathcal{U}(0,1)$ and
$\nu\sim \Gamma(a_\nu,\, b_\nu)$, $a_\nu, b_\nu>0$,
complete prior specification.

We now highlight some of the implications of the SICS prior on the overall graph structure and on the corresponding MCMC algorithm.
In the standard GGM framework, moves on the posterior space of graphs usually involve a single edge and consequently, when updating the graph, we only need to integrate out one element of the Cholesky decomposition of the precision matrix (assumed to have a $G$-Wishart prior) leading to efficient computation \citep{vandenBoom2022}. In such a context, updating more than an edge at a time is extremely challenging. On the other hand, under the SICS framework, change of block membership for a single node can affect a number of edges in the graph:
\begin{itemize}
\item The node joins a new block, and forms connections with every node already present in that block, thus a number of edges are introduced in the graph.
\item The node is removed from the current block (which is represented by a clique). Thus, as many edges as the number of nodes left in that block are removed. Some of these edges might be readded as there is a positive probability $\rho$ of connection between nodes in different communities.
\end{itemize}
From a computational point of view,
the structure in SICS poses a challenge to posterior inference using MCMC, due to the multiple edge change. Updates of $z$ conditional on $G$ are not possible.
Instead, we devise MCMC steps to update $z$ and $G$ jointly, in addition to updating $G$ conditional on $z$.
Section~\ref{sec:sics_mcmc} of Supplementary Material \add{\citep{vandenBoom2022_supp}} details the MCMC.

The SICS prior is a limiting case of the stochastic blockmodel, obtained taking the limit for $\beta^\star_k\to \infty$ in the model described in Section~\ref{sec:sbm}. In the limit, blocks become cliques with probability one.
\citet{Palla2012} consider a latent factor model that is, conditionally on the factor loadings, equivalent to a GGM with the SICS prior where $\rho = 0$. That is,
there is an edge between two nodes if and only if they belong to the same block.

\subsection{Multiple Graphs}
\label{sec:multi_sbm}

In applications, it is common that data (i.e.~the rows of the observation matrix $Y$) are naturally clustered due to experimental conditions.
For instance, $Y$ could represent gene expression measurements with different groups of rows corresponding to different cancer types.
One way to deal with such heterogeneity is a multiple graphical model \citep[e.g.][]{Peterson2015,
Ma2016,Mitra2016,Tan2017}, where each graph corresponds to an experimental condition (e.g.\ case/control status).

Consider multiple graphs $G_x=(V, E_x)$
and associated data $(n_x, Y_x)$ for $x=1,\dots,q$.
Here, $x$ indexes the groups such that we have $n_x$ observations in $Y_x$, with each group characterised by its own graph and data-generating process $p(Y_x\mid G_x)$.
We model the graphs $G_1,\dots,G_q$ jointly
through the specification of a prior $p(G_1,\dots,G_q)$. The goal is to identify common patterns, as well idiosyncratic edge/block structures.
We now introduce the multiple graph extension of the degree-corrected stochastic blockmodel, described in Section~\ref{sec:sbm}.
For each graph $G_x$, we have $\Pr\{(i,j) \in E_x \}=\Phi(\mu_{xij})$ and 
\eqref{eq:mu} becomes
\begin{equation*}
\mu_{xij} = \theta_i + \theta_j + \beta_{xij}\,
\mathds{1}
[z_{xi}=z_{xj}=k]
\end{equation*}
where $z_{xi}$ is the allocation variable for group~$x$, $x = 1, 
\dots, q$, and node $i\in V$. Similarly to Section~\ref{sec:sbm}, we introduce auxiliary variable $\beta_{xi}\mid H \stackrel{\text{i.i.d.}}{\sim} H$ for $i\in V$,
marginally for each $x$. Then, $\beta_{xij} = \beta_{xi} =\beta_{xj}$ when $i,j$ belong to the same community under condition $x$. The other parameters of the blockmodel are shared across graphs and have priors as specified in Section~\ref{sec:sbm}. Thus, marginally for each $x$, we recover the blockmodel of Section~\ref{sec:sbm}.

We treat one group (and so graph) as baseline
and the other graphs as offset from the baseline group. For ease of notation, we set $x=1$ as baseline group
and, for clarity, corresponding parameters by a subscript `$b$'.

There is a vast literature on \add{D}ependent Dirichlet processes \add{\citep[DDPs, see, e.g.,][]{MacEachern1999,DeIorio2004,Muller2004,Camerlenghi2019,Quintana2022}}, where the goal is to cluster subjects based also on group information.
These tools are not directly applicable to our context as we are actually clustering variables (i.e.~nodes on the graph) observed on $n_x$ subjects under each of $q$ experimental conditions (groups).
Since we are assuming that $\beta_{xi} \mid H\stackrel{\text{i.i.d.}}{\sim} H$ marginally for each group~$x$, the same node~$i$ under different groups can either belong to a different cluster (block) or to the same.
In the multiple graph context, it is then desirable to have
$\mathrm{Pr}(z_{bi}=z_{xi}) > \mathrm{Pr}(z_{bi}=z_{xj})$
for $i\ne j$ and $x\geq 2$
to reflect that
node $i$ in $G_b$ and in $G_x$ correspond to the same variable and to encourage sharing of large-scale structures across graphs.
\add{In a multiple networks context,  \citet{Reyes2016} require the same property.}
See also the discussion on identifiability for multiple blockmodels in Section~2.2 of \citet{Matias2016}
\add{and arguments for the related concept of separate exchangeability in \citet{Lin2021}.}
On the other hand,
DDP models typically assume exchangeability within each group~$x$ across subjects which implies that $\mathrm{Pr}(z_{xi} = k) = \mathrm{Pr}(z_{xj} =k )$ a priori.
We thus consider the following set-up, where the block structures $\{z_{xi}\}_{i=1}^p$, $x\geq 2$, are estimated as offsets from the baseline $\{z_{bi}\}_{i=1}^p$.

Let $\beta_{bi} \mid H \stackrel{\text{i.i.d.}}{\sim} H$ for $i\in V$ and $H\sim\mathrm{DP}(\nu, H_0)$
where $H_0 = \mathcal{N}(0, s^2_\beta)$.
Then, set $\beta_{xi} = \beta_{bi}$ with probability $\gamma\in(0, 1)$
and $\beta_{xi}\mid H \sim H$ with probability $1 - \gamma$,
independently for $i\in V$ and $x=2,\dots,q$.
\add{We note that our construction is invariant to any relabelling of the blocks. This is due to the fact that 
the distribution of  $\beta_{xi}$ is a mixture of a point mass at $\beta_{bi}$ and $H$:
\[
    \beta_{xi}\mid  \beta_{bi},H \sim \gamma \delta_{\beta_{bi}} +(1-\gamma) H
\]
where $H$ defines the overall 
 block structure (across multiple graphs). Hence, our prior specification allows keeping the same block labels across multiple graphs, while biasing the probability that  a node belongs to the same block across conditions.
Moreover, 
the implied dependence across block structures
closely resembles the type of  dependence across partitions described in
\citet{Page2022}. In a dynamic clustering framework, they  avoid the use of cluster labels by specifying a prior directly on random partitions, inflating the probability of belonging to the same cluster across time.}

Posterior computations are greatly simplified by the introduction of binary ``genealogical indicators'' $g_{xi}\in\{0,1\}$
such that $\beta_{xi} = \beta_{bi}$ if $g_{xi}= 1$
and $\beta_{xi} \mid H \sim H$ if $g_{xi}= 0$.
Note that even in the case $g_{xi}= 0$, there is a positive probability that $\beta_{xi} = \beta_{bi}$ due to the discrete nature of $H$. This implies that the probability of $\beta_{xi} = \beta_{bi}$, $x\geq 2$, is greater than $\mathrm{Pr}(g_{xi} = 1) = \gamma$ a priori conditionally on $H$.
Section~\ref{sec:multi_mcmc} of Supplementary Material \add{\citep{vandenBoom2022_supp}} details an MCMC algorithm for inference which involves a joint update for $(g_{xi}, \beta_{xi})$.
The prior dependence among the $z_{1i},\dots,z_{qi}$ enables learning of block structure both within and across graphs. The indicators $g_{xi}$ capture the extent to which structure in $G_x$ is shared with $G_b$. At the same time, the cluster indicators $\{z_{xi}\}_{i=1}^p$ capture the within-graph block structure. 
Thus, the proposed prior construction allows for borrowing of large-scale information across graphs, as well as the detection of graph-specific blocks.

We want to highlight that our construction differs from the hierarchical Dirichlet process \citep[HDP,][]{Teh2006}, as assuming an HDP-type prior would imply that $\beta_{xi}\mid H_x \stackrel{\text{ind.}}{\sim} H_x$, $H_x\mid H \stackrel{\text{i.i.d.}}{\sim} \mathrm{DP}(\cdot,H)$, $H\sim \mathrm{DP}(\nu, H_0)$. This means that the $\beta_{xi}$ have a positive probability to be equal across group~$x$ (and obviously node $i$), but the same node would not have higher probability to belong to the same block across groups.

There are proposals in the graphical model literature
where a graph is considered baseline
\citep[e.g.][]{Mukherjee2008,
Telesca2012,Mitra2016,Tan2017},
but their focus is on differences in individual edges instead of blocks.
Moreover, \citet{Paul2020}
consider blockmodels with multiple graphs
in a frequentist framework where the number of blocks is known and assume a hierarchical structure for the block memberships under each experimental condition, linking block membership to an unknown baseline membership. \citet{Reyes2016} and \citet{Stanley2016} induce dependence among graphs by assuming that they either share the same block structure or have unrelated block structures, which leads to a less flexible modelling tool than our approach.
\add{\citet{Amini2021} use an HDP to link block structure across graphs which presents the limitation described in the previous paragraph.}
\citet{Ma2016} assume that multiple graphs share the same known block structure and only edges between and across blocks might differ across experimental conditions. Edges are estimated using regularised nodewise regression \citep{Zhou2011}, instead of working directly on graph space.
In a different context, previous work on hidden Markov models \citep{Fox2008}, including time-varying blockmodels \citep{Ishiguro2010,Fan2015}, involves a similar dependence for cluster indicators across time, but this dependence is induced through a more involved construction with a ``spiked'' base measure for the DP on the transition probability vector of the Markov chain.

In our construction, the distribution of $\{z_{xi}\}_{i=1}^p$, $x\geq 2$, is defined conditionally on $\{z_{bi}\}_{i=1}^p$
such that $\{ g_{xi}\}_{i=1}^p$ captures differences from $\{z_{bi}\}_{i=1}^p$.
Alternative dependence structures could be easily considered within our framework. For instance, instead of setting a group as baseline, we could specify a latent block structure $\{z_{0i}\}_{i=1}^p$
and then define $\{z_{xi}\}_{i=1}^p$, $x\geq 1$, as deviations from $\{z_{0i}\}_{i=1}^p$, for which a prior process needs to be specified (e.g.\ simply assume $H$ as prior).
Finally, the multiple graph set-up can be straightforwardly extended to the SICS prior from Section~\ref{sec:sics}.

\section{Testing for Large-Scale Structure}
\label{sec:test}

In this section, we describe a strategy to test if there is any block structure in an individual graph. Although the description below will only involve one graph for ease of explanation, the same techniques can be employed to test for the presence of structure in multiple graphs.

In model \eqref{eq:mu}, block structure is represented through indicator vector $z$. Thus, testing for presence of block structure is equivalent to testing whether $K \geq 2$. In the Bayesian paradigm, we can use Bayes factors and here we describe a computational method for their evaluation based on the Savage-Dickey ratio \citep{Dickey1971}, as they are not available analytically.
Consider a prespecified block structure $z^\star$.
For instance, $z^\star$ can consist of a single ($K^\star = 1$) block, i.e.~no large-scale structure,
such that the test assesses the evidence for any block structure.
Our method will often be computationally infeasible for other choices of $z^\star$ as discussed later but the idea applies to any $z^\star$ in principle.

The computation of Bayes factors for DP-based models 
has received attention in the literature though
with some drawbacks:
the method from \citet{Basu2003}
requires an extra MCMC run with $z$ fixed to $z^\star$
and the use of sequential importance sampling,
resulting in an involved strategy, not easily
integrated into an existing MCMC implementation.
\citet{Legramanti2022_1} evaluate the marginal likelihood of each model using the harmonic mean approach
\add{\citep{Newton1994,Raftery2007}},
which can be unstable or slow to converge. Application of their method in our (and other's) context would benefit from the direct evaluation of $p(Y\mid z)$
which is not available in closed form. Without an analytical form for $p(Y\mid z)$, the harmonic mean approach requires an extra MCMC run with $z$ fixed to $z^\star$ to approximate $p(Y\mid z^\star)$.
Indeed, one of the main advantages of our method is that Bayes factors can be evaluated directly from the MCMC output for the 
model of interest.

More in details, the Bayes factor of the relative evidence of
$z = z^\star$ (model $\mathcal{M}^\star$) over $z\sim p(z)$ (model $\mathcal{M}$) is
\begin{equation} \label{eq:BF}  
    B = \frac{p(Y\mid \mathcal{M}^\star)}{p(Y \mid \mathcal{M})} = \frac{p(Y\mid z^\star)}{p(Y)}
    = \frac{p(Y,\, z^\star)}{p(Y)\, p(z^\star)}
    = \frac{p(z^\star\mid Y)}{p(z^\star)}
\end{equation}
where the last ratio is the Savage-Dickey ratio.
Note that the second equality uses the property that the prior
on all remaining parameters in the model such as $\beta_i$ are the same under $\mathcal{M}$ and $\mathcal{M}^\star$ in such a way that we recover the same model specification as $\mathcal{M}^\star$ when $z=z^\star$ in $\mathcal{M}$.
Now, an estimate $\widehat{B}$ for
$B$ is obtained by plugging in the usual (in terms of sample frequency) estimate of $p(z^\star\mid Y)$
derived from the MCMC chain
while $p(z^\star)$ is readily computed by numerical quadrature: $p(z) = \int p(z\mid \nu)\, p(\nu)\, d\nu$
where
$p(\nu) = \Gamma(\nu\mid a_\nu, b_\nu)$
and \citep[e.g.][]{Legramanti2022}
\[
    p(z\mid \nu) =
    \frac{\nu^K \prod_{k=1}^K (|\{i\mid z_i = k\}| - 1)!}{ \prod_{i=0}^{p-1} (\nu + i)}.
\]

This scheme can be employed to compute $\widehat{B}$
from an MCMC chain as long as
$p(z^\star)$
is not too small.
In that case,
reliably estimating $p(z^\star\mid Y)$
might be hard as the MCMC chain could visit $z^\star$ only rarely after convergence.
Furthermore,
$p(z^\star)$ will often be too small if $z^\star$ corresponds to multiple blocks
due to the combinatorially many ways to assign $p$ nodes to $K^\star\geq 2$ blocks, but it usually assumes reasonable values for $z^\star$ corresponding to absence of block structure ($K^\star = 1$),
which refers to the conventional null hypothesis of no structure (i.e.~``no effect'').
For instance, we test for $K^\star = 1$ in the examples considered in this work.
We remark that the methods from \citet{Basu2003} and \citet{Legramanti2022_1}
do not have such limitation for small $p(z^\star)$, but in general require additional MCMC runs.
Note that $p(z^\star\mid Y)$ being estimated as (close to) zero is not problematic, but leads to an accurate estimate of $B\approx 0$ (as long as $p(z^\star)$ is sufficiently far from zero). In Section~\ref{sec:karate_BF} of Supplementary Material \add{\citep{vandenBoom2022_supp}},
we show empirically that the proposed Bayes factor estimation converges faster than the harmonic mean approach.

\section{Simulation Studies}
\label{sec:simul}

We demonstrate the performance of our approach in two simulation scenarios.
For all empirical results,
we set the hyperparameters as $\gamma=0.5$, $s_\beta^2 = s_\theta^2 = 1$,
$a_\nu=b_\nu=a_\alpha=b_\alpha=2$,
$\delta=3$ and $D=I_p$
unless otherwise stated.
\add{See Section~\ref{sec:overview} of Supplementary Material \add{\citep{vandenBoom2022_supp}} for an overview of the models.}

\subsection{Karate Club Network}
\label{sec:karate}

We investigate the importance of uncertainty propagation when learning community structure in a graphical model.
As true underlying graph $G$, we consider the karate club network \citep{Zachary1977} which \citet{Tan2019} analyse using the degree-corrected stochastic blockmodel of Section~\ref{sec:sbm}.
The network's $p=34$ nodes correspond to members of a karate club while
its 78 edges signify friendships between members.
Conditionally on $G$,
we sample a precision matrix $\Omega\mid G \sim \mathcal{W}_G(\delta, D)$.
The $n$ rows of $Y$
are sampled
according to the GGM in Section~\ref{sec:ggm} independently from $\mathcal{N}(0_{p\times 1},\, \Omega^{-1})$.
Finally, we fit the model from Section~\ref{sec:sbm} using 6000 MCMC iterations, discarding the first 1000 as burn-in.
In this case, we set $a_\nu=b_\nu=a_\alpha=b_\alpha=5$ as in \citet{Tan2019} for a fair comparison.

\begin{figure}[tbp]
\centering
\includegraphics[width=\textwidth]{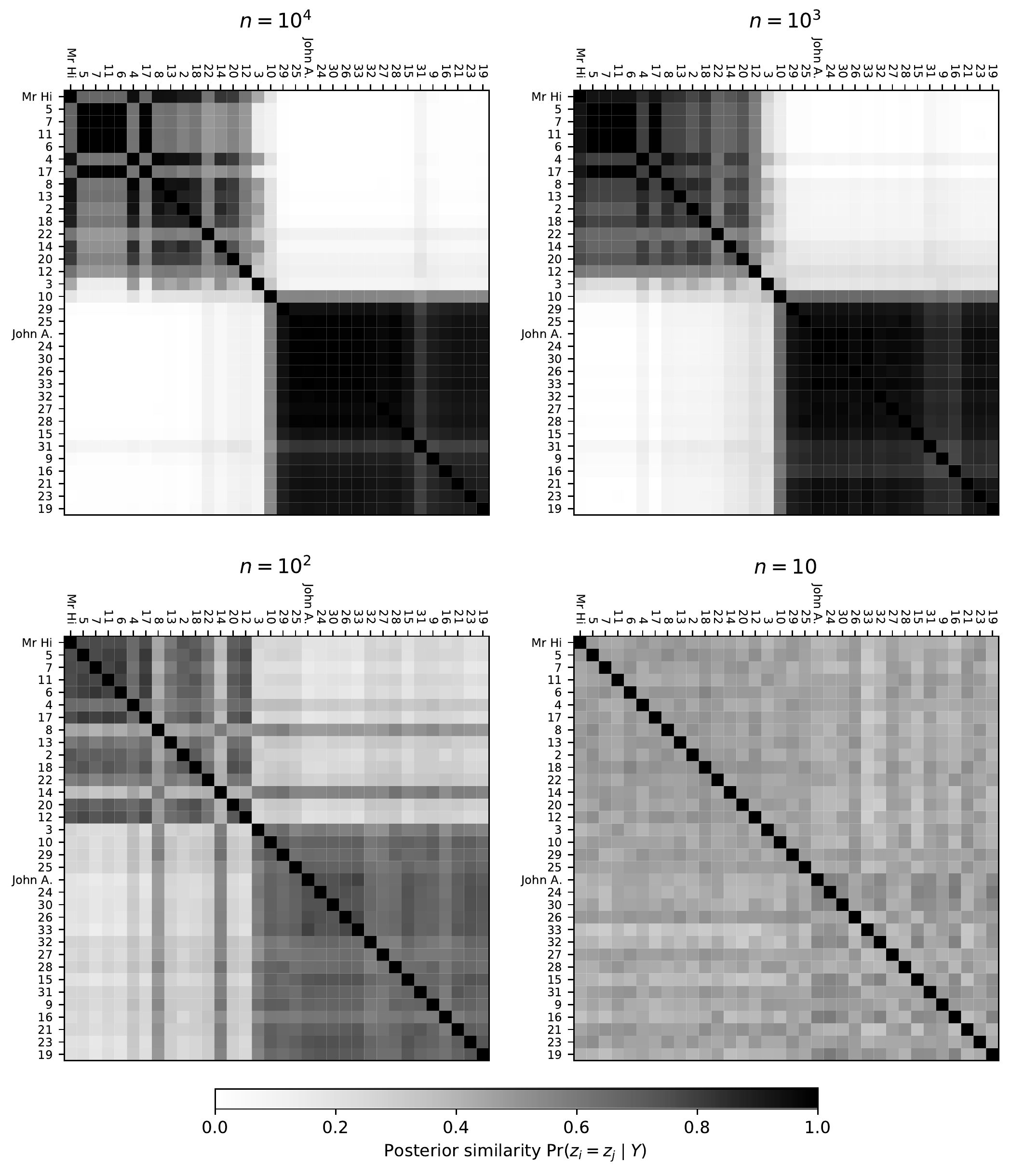}
\caption{ 
Karate club network:
posterior similarity matrices for the simulation studies. The nodes are ordered as in \citet{Tan2019}.
}
\label{fig:karate}
\end{figure}

We repeat the simulation for $n=10^4, 10^3, 10^2, 10$ while keeping $\Omega$ the same and present the estimated community structure in Figure~\ref{fig:karate}.
For $n=10^4$, the results are very close to those in \citet{Tan2019} where the underlying network is known,
with the two main blocks corresponding to the karate instructor Mr Hi and the club's president John A.
The increased uncertainty in the estimation of $G$ for smaller values of $n$ obviously affects inference on the block structure,
with too little information present in the data with only $n=10$ observations to recover the two main blocks. This is also reflected in
the estimate of the Bayes factor comparing the model with no block structure vs the model with $z\sim p(z)$: $\widehat{B} = 0$ for $n=10^4, 10^3$, $\widehat{B} = 0.28$ for $n=10^2$ and $\widehat{B}=0.40$ for $n=10$.
These results show that uncertainty in graph estimation can have a major impact on community estimation and this uncertainty should not be ignored as is often done in applications where a two-step approach is adopted (first graph estimation and then blocks).

\subsection{Block Structure Recovery}
\label{sec:recovery}

We now investigate how accurately the proposed methodology can recover block structure.
We assign $p=20$ nodes to
$K$ clusters
by sampling $z_i$ with replacement from
$\{1,\dots,K\}$ for $i\in V$.
Then, we generate a graph $G$ according to the SICS prior from Section~\ref{sec:sics} with between-block edge inclusion probability $\rho=0.2$.
Data~$Y$ corresponding to $G$ are sampled as in Section~\ref{sec:karate}.
We consider
the following scenarios:
$K = 4$ for $n = 20,100,500,1000$
and
$n=500$ for $K = 2,3,4,5$.
The performance of the algorithms is assessed over 50 replicates for each scenario.

We fit both models from
Sections~\ref{sec:sbm}
and
\ref{sec:sics},
as well as the model by \citet{Sun2014} (see Section~\ref{sec:sun} of Supplementary Material \add{\citep{vandenBoom2022_supp}} for a description) for comparison. We run the MCMC for 1000 iterations after a 
burn-in of 500 for the
for the stochastic blockmodel and
the model by \citet{Sun2014}
while we record 
5000 iterations after a burn-in of 1000 for the SICS model to account for the slower convergence and mixing of its MCMC.

The cluster allocation vector $z$ informs the block structure.
As point estimate for $z$, we report the configuration that minimises the posterior expectation of \citeauthor{Binder1978}'s \add{\citeyearpar{Binder1978}} loss function
under equal misclassification costs, which is a common choice in the applied Bayesian nonparametrics literature \citep{Lau2007}.
See Appendix~B of \citet{Argiento2014} for computational details. Briefly,
this expectation of the loss
measures
the difference for all possible pairs of nodes between the posterior probability of co-clustering and
the estimated cluster allocation.
Following \citet{Sun2014}, we use the \add{\citet{Rand1971}} index
to quantify the difference between the true allocation
and its \citeauthor{Binder1978} point estimate. A \citeauthor{Rand1971} index of one corresponds to a perfect match while a lower value indicates worse block structure recovery.

\begin{figure}[tbp]
\centering
\includegraphics[width=\textwidth]{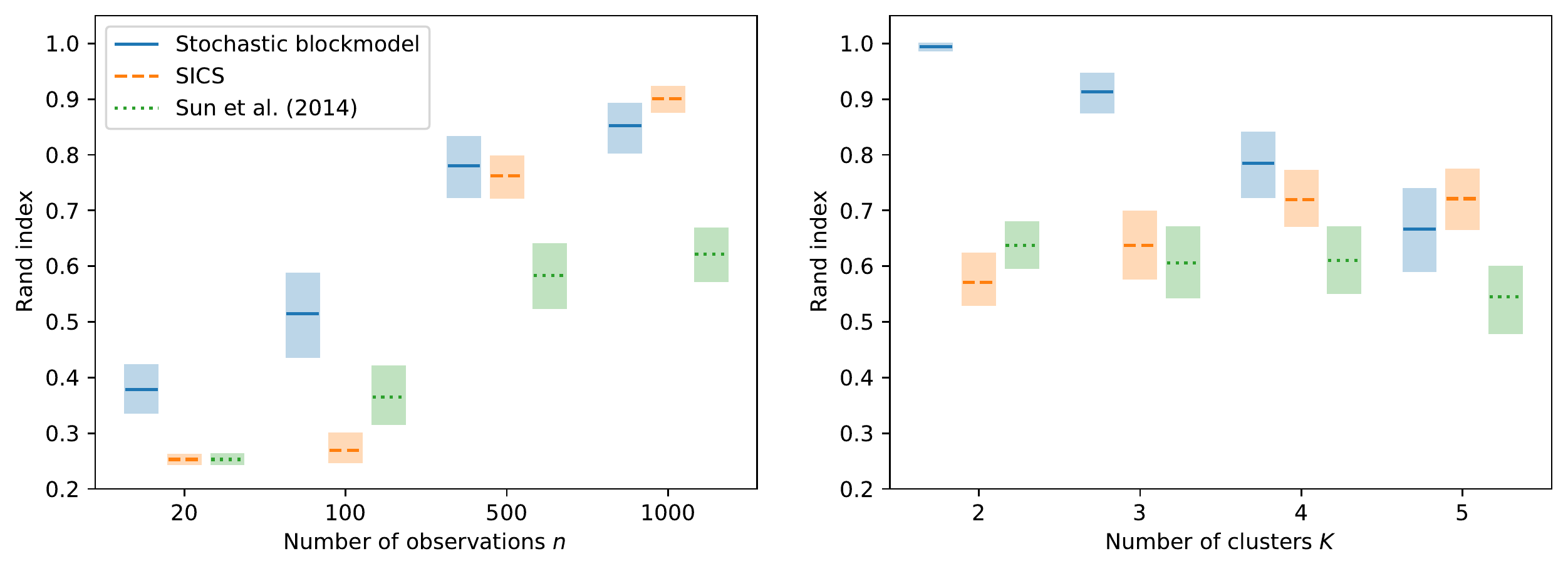}
\caption{
Block structure recovery:
\citeauthor{Rand1971} index versus the number of observations (left) and the number of clusters (right).
The lines represent means over the 50 replicates for the stochastic blockmodel, SICS and the model by \citet{Sun2014}.
The shaded areas are 95\% bootstrapped confidence intervals.
}
\label{fig:recovery}
\end{figure}

Figure~\ref{fig:recovery}
shows that the proposed methodology recovers the block structure comparably to or substantially more accurately than the model by \citet{Sun2014}.
The superior performance of the stochastic blockmodel over SICS, when occurring, is most likely due to the fact that the SICS model imposes more stringent assumptions on the correlation structure of the data, which might not be captured with small sample sizes (left panel of Figure~\ref{fig:recovery}).
This is in line with power considerations for detecting correlation in a frequentist framework, with large sample size
usually required, especially for partial correlations \citep[see, e.g.,][]{Castelo2006,Knudson2014}.
Secondly, the SICS structure is more easily recovered when fewer nodes belong to a block
(right panel of Figure~\ref{fig:recovery}) as this relaxes the assumption on the overall dependence structure among the random variables.
Finally, posterior inference for the stochastic blockmodel is performed through an exact MCMC \citep{vandenBoom2022}
while, for the SICS, we employ a Laplace approximation for the graph likelihood
to update $z$ and $G$ jointly. See Section~\ref{sec:mcmc} of Supplementary Material \add{\citep{vandenBoom2022_supp}}.

\section{Applications}
\label{sec:applications}

We apply the proposed models to two real data sets.
We discuss MCMC mixing and convergence
in Section~\ref{sec:trace_plots} of Supplementary Material \add{\citep{vandenBoom2022_supp}}.

\subsection{Mutual Fund Data}
\label{sec:fund}

We consider data on monthly returns of $p=59$ mutual funds described in
\citet{Scott2008}.
The funds are divided into four types by the sectors they invest in
with 13 funds investing in U.S.\ bonds, 30 in U.S.\ stocks, 7 in both U.S.\ stocks and bonds, and 9 in international stocks.
The data contain observations on $n=86$ months.
Here, we ignore the dependence of the returns across time
and focus on the dependence between funds as in \citet{Scott2008} and \citet{Marlin2009b}. Note that time dependence could be easily incorporated through a mean term. The returns are quantile-normalised so that they marginally follow a standard Gaussian distribution.
We fit both the degree-corrected stochastic blockmodel from Section~\ref{sec:sbm}
and SICS from Section~\ref{sec:sics}.
We run the MCMC chain for 15000 iterations discarding the first 5000 as burn-in for the stochastic blockmodel, and for 110000 iterations discarding the first 10000 for SICS.

\begin{figure}[tb]
\centering
\includegraphics[width=\textwidth]{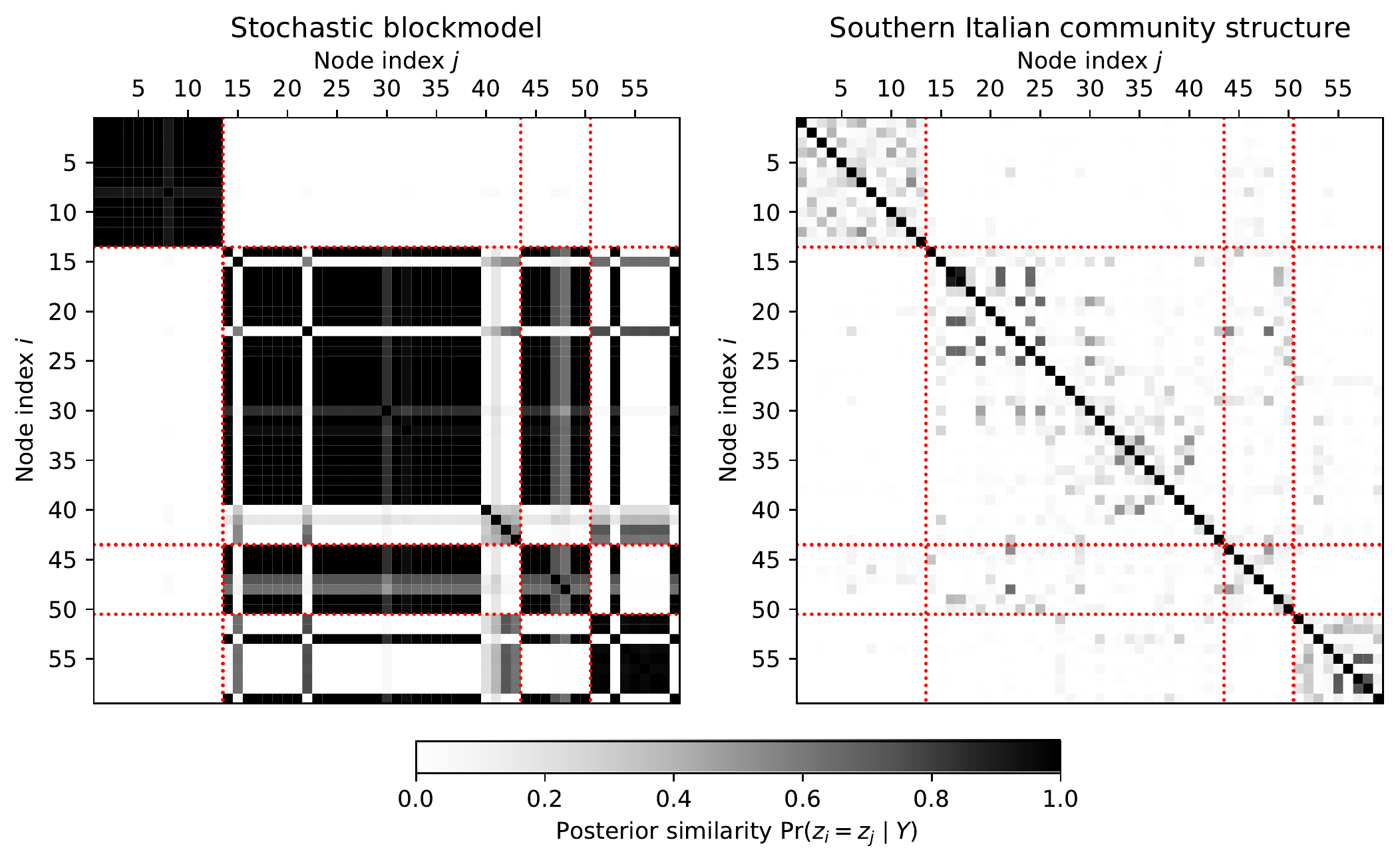}
\caption{
Mutual fund data: posterior similarity matrices.
The red dotted lines demarcate the fund types.
}
\label{fig:fund}
\end{figure}

\begin{figure}[tbp]
\centering
\includegraphics[width=\textwidth]{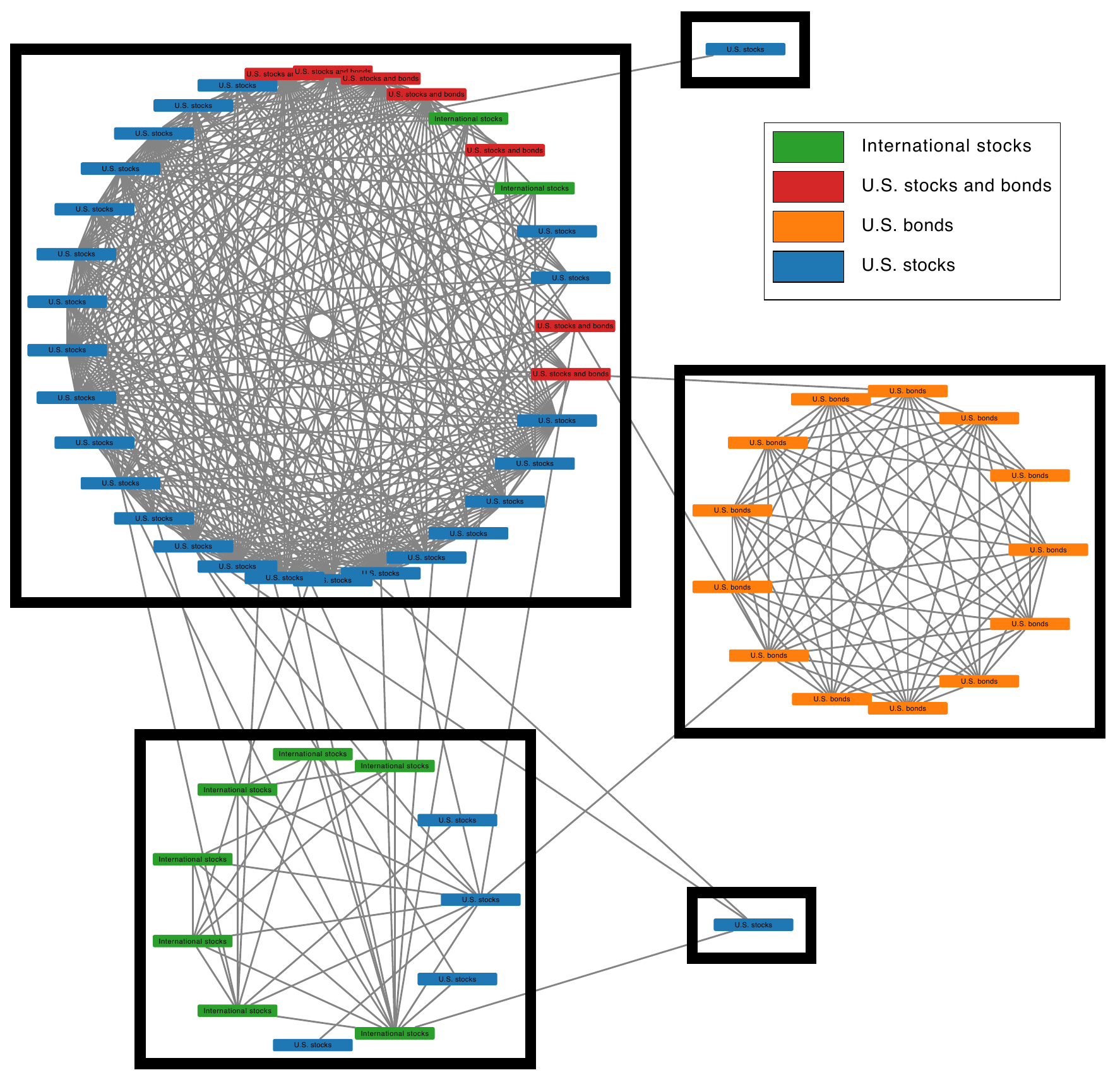}
\caption{Mutual fund data:
median probability graph \citep{Barbieri2004},
which consists of edges with posterior inclusion probability greater than $0.5$,
from the stochastic blockmodel. Colour of nodes refers to fund type. Boxes group nodes belonging to blocks estimated by minimising
\citeauthor{Binder1978}'s loss function.
}
\label{fig:fund_graph}
\end{figure}

The stochastic blockmodel identifies clear blocks of funds per Figures~\ref{fig:fund} (left panel) and \ref{fig:fund_graph}.
Specifically, the U.S.\ bonds, U.S.\ stocks and bonds, and international stocks funds are each blocked together without overlap between these fund types
except for
two international stocks funds that are grouped with the U.S.\ stocks and bonds.
The other funds, which invest in U.S.\ stocks,
are mostly blocked with the U.S.\ stocks and bonds, or with the international stocks but not with the U.S.\ bonds.
These results are intuitive as funds with a mixture of U.S.\ stocks and bonds make investments which overlap with funds with only U.S.\ stocks,
and correlation between the returns of U.S.\ and international stocks is likely.
This identified block structure is notably more in line with the fund types than the blocking results presented in \citet{Marlin2009b} obtained by shrinkage estimation and optimisation, where only a clear separation of the U.S.\ bonds funds from the others
is detected.

The SICS prior leads to a
large number of blocks with a posterior mode at $K=24$ blocks.
A larger number of blocks with SICS than with the stochastic blockmodel is expected as SICS' definition of a block as a clique is more stringent such that larger blocks are less likely to appear.
The large-scale pattern of the similarity matrix for SICS is still similar to that of the stochastic blockmodel in Figure~\ref{fig:fund},
though with much lower values for $\mathrm{Pr}(z_i=z_j\mid Y)$.
Still,
the posterior fit indicates strong evidence for the presence of a block structure as the Bayes factor in favour of absence of block structure is $\widehat{B} = 0$.

Generally,
the posterior inference contains information on whether the stochastic blockmodel or SICS is the most appropriate model for the data at hand.
We assess this by computing the Bayes factor of the stochastic blockmodel versus SICS using the harmonic mean approach.
This results in a log Bayes factor of 235 indicating strong 
evidence that the stochastic blockmodel fits the data better than SICS.

\subsection{Gene Expression Data}
\label{sec:gene}

As a second application, we consider gene expression levels,
the interactions between which are often represented as networks.
An important concept in the gene network literature is that of module, which is a densely connected subgraph.
Thus, learning the block structure of a graph allows module detection.
Typically, a two-step approach is adopted: first the graph is estimated from the gene expressions, and then the modules are derived from the graph estimate \citep[see, e.g.,][]{Zhang2018}, which underestimates uncertainty and often leads to false positives.

We analyse data on gene expressions
from $n_1=590$ breast cancer tissue samples and $n_2=561$ ovarian cancer samples
from
The Cancer Genome Atlas.
We focus on $p=44$ genes
identified by \citet{Zhang2018} as
spread across four estimated modules (Modules~6, 14, 36 and 39 in Table 2 of the cited paper)
which are highly enriched in terms of Gene Ontology \citep[GO,][]{Ashburner2000} annotations.
For each cancer, the gene expressions are quantile-normalised to marginally follow a standard Gaussian distribution.
We apply the proposed multiple graph methodology from Section~\ref{sec:multi_sbm} with $q=2$ separate groups, corresponding to the two different cancers.
We run the MCMC algorithm for 55000 iterations, discarding the first 5000 as burn-in.

\begin{figure}[tb]
\centering
\includegraphics[width=\textwidth]{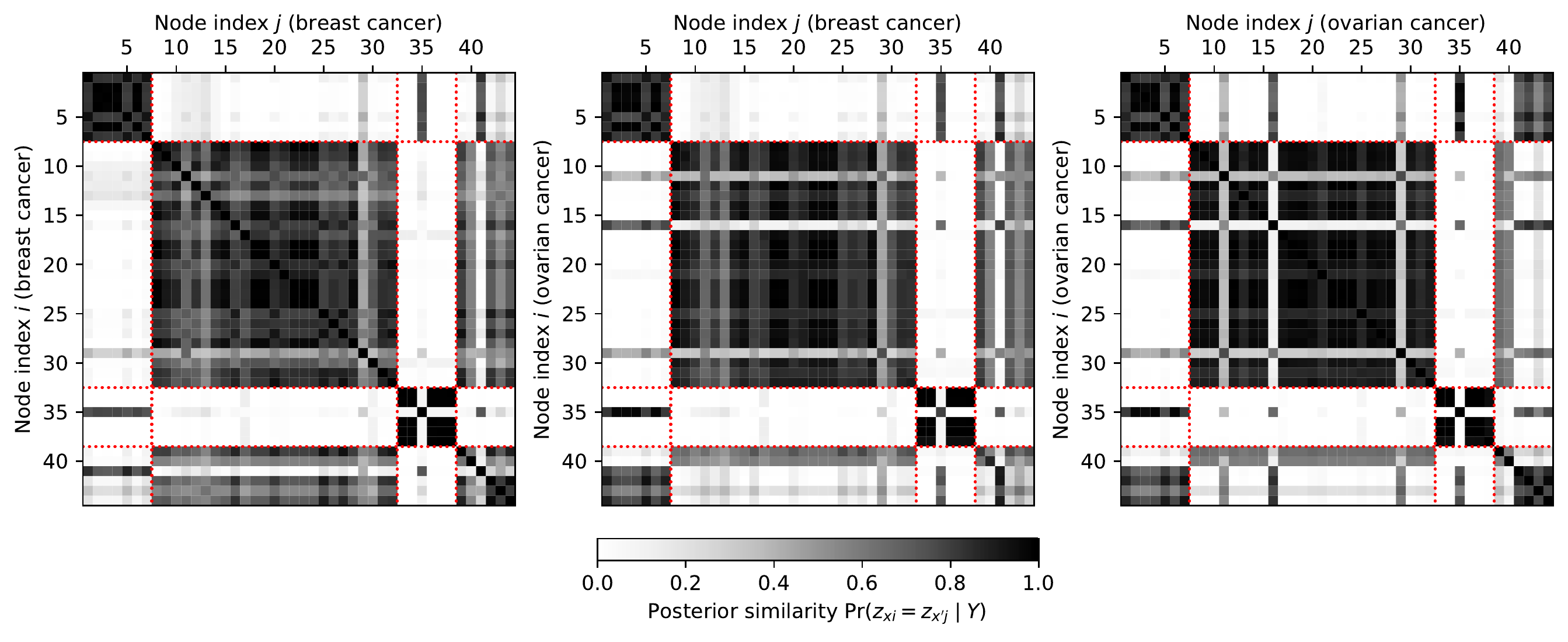}
\caption{
Gene expression data:
posterior similarity matrices.
The three matrices visualise the posterior probabilities
of
$z_{1i} = z_{1j}$,
$z_{1i} = z_{2j}$
and
$z_{2i} = z_{2j}$, respectively,
where $x=1$ corresponds to breast cancer and and $x=2$ to ovarian cancer.
The red dotted lines demarcate the modules identified by \citet{Zhang2018}.
}
\label{fig:gene}
\end{figure}

Posterior inference on block structure,
shown in Figure~\ref{fig:gene},
carries strong similarities with the modules identified by \citet{Zhang2018}.
For ease of discussion,
we refer to the modules from \citet{Zhang2018}
as Module~1 (comprising Nodes~1 through 7), 2 (Nodes~8 through 32), 3 (Nodes~33 through 38) and 4 (Nodes~39 through 44), and highlight them in Figure~\ref{fig:gene}.
The proposed methodology finds differences in block structure between breast and ovarian cancer (see middle panel of Figure~\ref{fig:gene})
as well as differences from \citeauthor{Zhang2018}'s modules, which are forced to be the same across both cancers by construction.
Across both breast and ovarian cancer, we find that Nodes~39 and 40 (GSTM3 and BCAR3 genes, respectively) are grouped with genes from Module~2, which has GO annotations relating to inflammatory response. For Breast cancer, we cluster Nodes~42 through 44 (GSTM1, GSTM2 and GSTM5) with Module~2 while we put them together with Module~1, which has GO annotations relating to pattern specification, for ovarian cancer.
Note that these three genes are paralogs of each other which suggests that they have similar function. Finally,
Nodes~35 and 41 (HOXB13 and GSTM4) are together with nodes in Module~1 for both cancers.
These results show the flexibility of the proposed model to capture differences
as well as commonalities
in large-scale dependence structure across multiple biological conditions.

\section{Discussion}
\label{sec:discussion}

In this work,
we combine advances from random graph theory with graphical models to obtain joint estimation of the graph and its large-scale structure.
The resulting graphical models are able to go beyond estimation of individual edges to provide inference on the community structure of the graph, while appropriately propagating uncertainty in the estimation.
We introduce a novel DDP prior process
tailored to the multiple graph setting
and propose a convenient computation of Bayes factors in partition models.
Advantages of the proposed approach include interpretability, flexibility (due to the nonparametric component) and wide applicability.
We focus on two different block structures: stochastic blockmodels and SICS. We note that the SICS prior is more suitable in applications where strong partial correlation between a small number of nodes is expected.

Alternative priors on the block structure could be considered.
For instance,
Gibbs-type priors \citep{Gnedin2005}
and
microclustering priors \citep{Betancourt2022}
are drop-in replacements for the respective DP terms used in \eqref{eq:dir}, and cover a wide range of partition priors
such as mixture \add{with random number of components} \citep{Miller2017,Geng2018,Argiento2022}.
In general, standard Bayesian nonparametric priors assume that the location parameters are i.i.d.\ draws from a base measure.
Note that, in our case, the location parameters, $\beta^\star_k$, correspond to the within-block interaction strength and they can be arbitrarily close or far given the prior assumptions.
On the other hand, in the context of mixture \add{with random number of components}, it is easier to introduce constraints on the locations, if the application or interpretability require it.
For instance, we might impose that the $\beta^\star_k$ vary significantly across blocks and, to that end, assume a
repulsive mixture prior \citep{Petralia2012}.

We focus on GGMs for convenience and because of their popularity. Our methodological contribution is however not constrained to this specific set-up and can be extended to work with other graphical models, e.g.\ to graphs with discrete or mixed type nodes.
Moreover,
our computational strategy to estimate
Bayes factors finds general applicability in the context of Bayesian nonparametric models to test the presence/absence of a partition structure.

\add{We conclude the discussion with few remarks on asymptotic properties of the proposed methodology.
Posterior contraction on the graph space does not imply contraction for the block structure and vice versa:
posterior contraction of the graph requires the number of observations $n\to\infty$.
See \citet{Lee2021} and \citet{Niu2021}, and references therein.
On the other hand,
asymptotic results for Bayesian learning  of block structure involve the number of nodes $p\to\infty$ \citep[e.g.][]{Geng2018,Gao2020,Jiang2021}.
For $n$ large, the graph might be estimated with high precision.
In that scenario, considerable posterior uncertainty about the block structure can remain as such uncertainty can be present even if the graph is observed.
Furthermore, for large $p$, block structure can be recovered accurately even if the estimates of the overall graph connection pattern are characterised by high uncertainty.}

\begin{supplement}
\stitle{Supplement}
\sdescription{\add{Overview of notation and models,} details of the MCMC algorithms, description of the model by \citet{Sun2014}, empirical results for the Bayes factor approach\add{,} and MCMC trace plots.}
\end{supplement}

\begin{supplement}
\stitle{Code}
\sdescription{Code for the empirical results is available at
\url{https://github.com/willemvandenboom/graph-substructures}.}
\end{supplement}

\bibliographystyle{ba_no_url}  
\bibliography{graph}

\begin{acks}[Acknowledgments]
The data used in Section~\ref{sec:gene}
are generated by The Cancer Genome Atlas Research Network: \url{https://www.cancer.gov/tcga}.
\end{acks}

\end{document}


\title{\bf Supplement to\\``Bayesian Learning of Graph Substructures''}
\author{\if0\blind Willem van den Boom, Maria De Iorio and Alexandros Beskos\fi}
\maketitle

\add{
\section{Overview of Notation and Models}
\label{sec:overview}

Here we provide an overview of the model notation (in Table~\ref{tab:overview_notation})
and model specification.
\begin{table}
\add{
\centering
\begin{tabular}{r|l}
     \textbf{Notation} & \textbf{Description} \\
     \hline
     $Y$ & Data matrix \\
     $\Omega$ & Precision matrix corresponding to graph $G$ \\
     $G = (V,E)$ & Graph \\
     $V$ & Set of nodes \\
     $E$ & Set of edges \\
     $\Phi(\cdot)$ & Cumulative distribution function of $\mathcal{N}(0,1)$ \\
     $z_i$ & Community that node $i$ belongs to \\
     $\beta_i$ & Interaction strength of community $z_i$ \\
     $\theta_i$ & Popularity parameter of node $i$ \\
     $H$ & Almost surely discrete random measure for the block structure \\
     $F$ & Almost surely discrete random measure for the popularity parameters \\
     $\nu$ & Concentration parameter of the DP for the block structure \\
    $\alpha$ & Concentration parameter of the DP for the popularity parameters \\
     $Y_x$ & Data matrix of group $x$ \\
     $G_x$ & Graph corresponding to group $x$ \\
     $\Omega_x$ & Precision matrix corresponding to graph $G_x$ \\
     $E_x$ & Set of edges of graph corresponding to group $x$ \\
     $z_{xi}$ & Community that node $i$ belongs to in group $x$ \\
     $\beta_{xi}$ & Interaction strength of community $z_{xi}$
\end{tabular}
\caption{Overview of model notation.}
\label{tab:overview_notation}
}
\end{table}
The likelihood for the single graph scenario is
\[
    Y\mid \Omega\sim\mathcal{MN}(0_{n\times p},\, I_n,\, \Omega^{-1})
\]
where
$\mathcal{MN}(0_{n\times p},\, I_n,\, \Omega^{-1})$
denotes a Matrix Gaussian distribution such that each row of $Y$ is distributed independently according to $\mathcal{N}(0_{p\times 1},\, \Omega^{-1})$. The density of a Matrix Gaussian distribution is given by
\[
    \mathcal{MN}(Y\mid M,\, U,\, \Omega^{-1})
    = (2\pi)^{-np/2} |\Omega|^{n/2} |U|^{-p/2}\exp\left[
        -\frac{1}{2}\mathrm{tr}\{
            \Omega(Y - M)^\top U^{-1} (Y - M)
        \}
    \right]
\]
for $Y\in\mathbb{R}^{n\times p}$
where $\mathrm{tr}(\cdot)$ denotes the matrix trace.
The prior on the precision matrix is
\[
    \Omega\mid G\sim \mathcal{W}_G(\delta, D)
\]
For the multiple graph scenario, we analogously have
\begin{align*}
    Y_x\mid \Omega_x &\overset{\text{ind.}}{\sim}\mathcal{MN}(0_{n\times p},\, I_n,\, \Omega_x^{-1}), \quad x = 1,\dots,q \\
    \Omega_x\mid G_x &\overset{\text{ind.}}{\sim} \mathcal{W}_{G_x}(\delta, D)
\end{align*}

The prior distributions on graphs are as follows.

\subsection{Degree-Corrected Stochastic Blockmodel}

\begin{alignat*}{2}
    \mathrm{Pr}\{(i,j)\in E\} &= \Phi(\mu_{ij}) &&\quad 1\leq i<j\leq p \\
    \mu_{ij} &= \theta_i + \theta_j + \beta_{ij}\,
\mathds{1}
[z_i=z_j] \\
    \beta_{ij} &= \beta_i = \beta_j &&\quad \text{if $\beta_i=\beta_j$} \\
    \beta_{i}\mid H &\overset{\text{i.i.d.}}{\sim} H
    &&\quad i\in V \\
    H &\sim \mathrm{DP}(\nu, H_0) \\
    H_0 &= \mathcal{N}(0, s^2_\beta) \\
    \theta_i\mid F &\stackrel{\text{i.i.d.}}{\sim} F \\
    F &\sim\mathrm{DP}(\alpha, F_0) \\
    F_0 &= \mathcal{N}(0, s^2_\theta) \\
    \nu &\sim \Gamma(a_\nu,\, b_\nu) \\
    \alpha &\sim \Gamma(a_\alpha,\, b_\alpha)
\end{alignat*}

\subsection{Southern Italian Community Structure}

\begin{alignat*}{2}
    \mathrm{Pr}\{(i,j)\in E\} &= \begin{cases}
        1 \quad &\text{if $z_i = z_j$} \\
        \rho \quad &\text{if $z_i \ne z_j$}
    \end{cases} &&\qquad 1\leq i<j\leq p \\
    z_i &\overset{\text{i.i.d.}}{\sim} \mathrm{CRP}(\nu) 
    &&\qquad \text{with $\mathrm{CRP}(\nu)$ a Chinese restaurant process}
    \\
    \rho&\sim\mathcal{U}(0,1) \\
    \nu &\sim \Gamma(a_\nu,\, b_\nu)
\end{alignat*}

\subsection{Multiple Graphs}
\label{sec:overview_multi}

\begin{alignat*}{2}
    \mathrm{Pr}\{(i,j)\in E_x\} &= \Phi(\mu_{xij}) &&\quad 1\leq i<j\leq p, \quad x=1,\dots,q \\
    \mu_{xij} &= \theta_i + \theta_j + \beta_{xij}\,
    \mathds{1}
    [z_{xi}=z_{xj}] \\
    \beta_{xij} &= \beta_{xi} = \beta_{xj} &&\quad \text{if $\beta_{xi}=\beta_{xj}$} \\
    \beta_{bi}\mid H &\overset{\text{i.i.d.}}{\sim} H
    &&\quad i\in V,\ \text{with $\beta_{bi} = \beta_{1i}$} \\
    \beta_{xi}\mid \beta_{bi}, H &\overset{\text{i.i.d.}}{\sim} \gamma\delta_{\beta_{bi}} + (1-\gamma) H
    &&\quad x=2,\dots,q,\ \text{with $\delta_{\beta_{bi}}$ a point mass at $\beta_{bi}$} \\
    H &\sim \mathrm{DP}(\nu, H_0) \\
    H_0 &= \mathcal{N}(0, s^2_\beta) \\
    \theta_i\mid F &\stackrel{\text{i.i.d.}}{\sim} F \\
    F &\sim\mathrm{DP}(\alpha, F_0) \\
    F_0 &= \mathcal{N}(0, s^2_\theta) \\
    \nu &\sim \Gamma(a_\nu,\, b_\nu) \\
    \alpha &\sim \Gamma(a_\alpha,\, b_\alpha)
\end{alignat*}

}

\section{Markov Chain Monte Carlo Algorithms}
\label{sec:mcmc}
This section discusses the Markov chain Monte Carlo (MCMC) algorithms used in the main text.

\subsection{Degree-Corrected Stochastic Blockmodel}

Algorithm~\ref{alg:sbm} details an MCMC step
to compute the joint posterior on $(G,\theta,\beta,z)$ with the degree-corrected stochastic blockmodel as prior on the graph space.
It uses the $G$-Wishart weighted proposal algorithm (WWA) of \citet{vandenBoom2022}
to update the graph $G$.
We use the latent variable update from \citet{Albert1993} for $\beta^\star$ and $\theta$.
The update of the block labels $z$ uses that \eqref{eq:dir} also applies to the Dirichlet process.
Define the cluster label $c_i$ for the popularity parameter $\theta$
analogously to $z_i$ for $\beta$,
and let $M$ denote the number of unique elements in $\theta$ such that $1\leq c_i\leq M$.
Then,
analogously to \eqref{eq:dir},
\begin{equation} \label{eq:dir_theta}
	\mathrm{Pr}(c_i=m\mid c_{-i}) =
	\begin{cases}
		\frac{n^\theta_{-i,k}}{p - 1 + \alpha}, &k=1,\dots,M^{-i} \\
		\frac{\alpha}{p - 1 + \alpha}, &m=M^{-i}+1
	\end{cases}
\end{equation}
where $n^\theta_{-i,k} = |\{1\leq j \leq p \mid c_j = k,\, j\ne i\}|$
and $M^{-i}$ is the number of unique elements in $c_{-i}=\{c_j\mid j\ne i\}$.
Also, denote the $M$-dimensional vector with the unique values of $\theta$ by $\theta^\star$.
The updates for the Dirichlet precision parameters $\nu$ and $\alpha$ follow \citet{Escobar1995}.

Algorithm~\ref{alg:sbm} is similar to Algorithm~1 of \citet{Tan2019}.
A difference is that we marginalise over the latent variables $\zeta_{ij}$ when updating the block labels $z_i$.

\begin{breakablealgorithm}
\caption{MCMC step for the degree-corrected stochastic blockmodel. \label{alg:sbm}}
\begin{enumerate}
	\item \label{step:G}
	Update $G\mid \mu,Y$ using WWA from \citet{vandenBoom2022}.
	\item
	For all $1\leq i<j\leq p$:
	\begin{enumerate}
		\item
		If $(i,j)\in E$,
		sample $\zeta_{ij}\sim \mathcal{N}_{(0,\infty)}(\mu_{ij}, 1)$.
		\item
		If $(i,j)\notin E$,
		sample $\zeta_{ij}\sim \mathcal{N}_{(-\infty, 0)}(\mu_{ij}, 1)$.
	\end{enumerate}
	\item
	For $k=1,\dots,K$,
	let $S^\beta_k = \{(i,j)\mid i < j,\, z_i = z_j = k\}$ and sample
	$\beta^\star_k \sim \mathcal{N}\{\sigma^2_k \sum_{(i,j)\in S^\beta_k} (\zeta_{ij} - \theta_i - \theta_j),\, \sigma^2_k\}$
	where $\sigma^2_k = (s_\beta^{-2} + |S^\beta_k|)^{-1}$.
	\item
	Let $\zeta_{ij}^{-\beta} = \zeta_{ij} - \mathds{1}
[z_i=z_j]\beta_i$ for $1\leq i<j\leq p$
and set $\zeta_{ji}^{-\beta} = \zeta_{ij}^{-\beta}$ for notational convenience.
	For $m=1,\dots,M$,
	let $S^\theta_m = \{(i,j)\mid i < j,\, c_i = c_j = m\}$ and sample
	$\theta^\star_m\sim \mathcal{N}[\sigma^2_m \{2\sum_{S^\theta_m} \zeta_{ij}^{-\beta} + \sum_{\{(i,j)\mid c_i = m \ne c_j\}} (\zeta_{ij}^{-\beta} - \theta_j)\}, \sigma^2_m]$
	where $\sigma^2_m = \{s_\theta^{-2} + 4|S^\theta_m| + n^\theta_m(p-n^\theta_m)\}^{-1}$
	with $n^\theta_m = |\{1\leq j \leq p \mid c_j = k\}|$.
	\item
	Update the block labels $z$. For $i=1,\dots,p$:
	\begin{enumerate}
	\item
	Set $K^{-i}$ equal to the number of unique values in $z_{-i}$
	and
	relabel $z_{-i}$ such that $1\leq z_j\leq K^{-i}$ for $j\ne i$.
	\item
	Denote the $\mu$ resulting from $z_i=k$ by $\mu^k$.
	Here, $\mu^{K^{-i}+1}$ does not involve the not yet specified $\beta^\star_{K^{-i}+1}$.
	The neighbourhood of node $i$ is $N_i = \{j\mid (i,j)\in E\}$.
	Sample $z_i$ according to
	\[
		p(z_i=k\mid \text{---})\propto
		\mathrm{Pr}(z_i=k\mid z_{-i})
		\prod_{j\in N_i} \Phi(\mu^k_{ij}) \prod_{j\notin N_i} \{ 1 - \Phi(\mu^k_{ij}) \}
	\]
	for $k=1,\dots,K^{-i}+1$
	where
	$\mathrm{Pr}(z_i=k\mid z_{-i})$
	is given by \eqref{eq:dir}.
	\item
	If $z_i \leq K^{-i}$, set $\beta_i = \beta^\star_{z_i}$ and $K=K^{-i}$.
	\item
	If $z_i = K^{-i} + 1$,
	sample $\beta_i\sim\mathcal{N}(0,s^2_\beta)$
	and set $K = K^{-i} + 1$.
	\end{enumerate}
	\item
	Update the block labels $c$. For $i=1,\dots,p$:
	\begin{enumerate}
	\item
	Set $M^{-i}$ equal to the number of unique values in $c_{-i}$
	and
	relabel $c_{-i}$ such that $1\leq c_j\leq M^{-i}$ for $j\ne i$.
	\item
	Sample $c_i$ according to
	\begin{multline*}
	p(c_i=m\mid \text{---})\propto
	\mathrm{Pr}(c_i=m\mid c_{-i}) \\
	\times
	\begin{cases}
		\exp\left\{
		    \theta^\star_m \sum_{j\ne i} (\zeta_{ij}^{-\beta} - \theta_j)
		    - \frac{p - 1}{2}(\theta^\star_m)^2
		\right\}, \quad &m=1,\dots,M^{-i} \\
		\frac{\sigma_c}{s_\theta} \exp\left( \frac{\mu_c^2}{2\sigma_c^2} \right),
		\quad &m = M^{-i} + 1
	\end{cases}
	\end{multline*}
	where
	$\mathrm{Pr}(c_i=m\mid c_{-i})$
	is given by \eqref{eq:dir_theta},
	$\sigma_c^2 = (p-1 + s_\theta^{-2})^{-1}$
	and
	$\mu_c = \sigma_c^2\sum_{j\ne i}(\zeta_{ij}^{-\beta} - \theta_j)$.
	\item
	If $c_i \leq M^{-i}$, set $\theta_i = \theta^\star_{c_i}$ and $M=M^{-i}$.
	\item
	If $c_i = M^{-i} + 1$,
	sample $\theta_i\sim\mathcal{N}(\mu_c, \sigma_c^2)$
	and set $M = M^{-i} + 1$.
	\end{enumerate}
	\item \label{step:nu}
	Sample $t_\nu\sim \mathrm{Beta}(\nu + 1,\, p)$.
	Then, sample $\nu$ from the mixture
	$\pi_\nu \Gamma(a_\nu + K,\, b_\nu - \log t_\nu) + (1-\pi_\nu) \Gamma(a_\nu + K - 1,\, b_\nu - \log t_\nu)$
	where $\pi_\nu$ is defined by
	\[
	    \frac{\pi_\nu}{1 - \pi_\nu} = \frac{a_\nu + K - 1}{p\,(b_\nu - \log t_\nu)}
	\]
	\item \label{step:alpha}
	Sample $t_\alpha\sim \mathrm{Beta}(\alpha + 1,\, p)$.
	Then, sample $\alpha$ from the mixture
	$\pi_\alpha \Gamma(a_\alpha + M,\, b_\alpha - \log t_\alpha) + (1-\pi_\alpha) \Gamma(a_\alpha + M - 1,\, b_\alpha - \log t_\alpha)$
	where $\pi_\alpha$ is defined by
	\[
	    \frac{\pi_\alpha}{1 - \pi_\alpha} = \frac{a_\alpha + M - 1}{p\,(b_\alpha - \log t_\alpha)}
	\]
\end{enumerate}
\end{breakablealgorithm}

\subsection{Southern Italian Community Structure}
\label{sec:sics_mcmc}

As mentioned in Section~\ref{sec:sics},
MCMC with the Southern Italian community structure as graph prior requires a joint update of $z$ and $G$.
A change in $z$ can force the addition of more than one edge in $G$.
Therefore, we cannot use WWA \citep{vandenBoom2022} used in Algorithm~\ref{alg:sbm}
as it is restricted to single edge updates.
Instead,
we evaluate $p(Y\mid G)$ in \eqref{eq:ggm_lik} directly by using the Laplace approximation, with a diagonal Hessian matrix, of the normalising constant $I_G(\delta, D)$ from \citet{Moghaddam2009}.
Then, we can readily evaluate Metropolis-Hastings updates that involve changing multiple edges in $G$.
Algorithm~\ref{alg:sics}
details the MCMC step.
Here, we only discuss the joint update of $z$ and $G$ in more detail (Step~\ref{step:joint}) as the other steps are more standard.

\begin{algorithm}
\caption{MCMC step for the Southern Italian community structure. \label{alg:sics}}
\begin{enumerate}
    \item
    Compute $n^\beta = \sum_k n^\beta_k(p - n^\beta_k) / 2$ where $n^\beta_k =
    |S^\beta_k| = |\{(i,j)\mid i < j,\, z_i = z_j = k\}|$.
    Sample $\rho$ from its full conditional
    $\mathrm{Beta}\{1 + |E| - n^\beta,\, 1 + p(p-1)/2 - |E|\}$\add{.}
	\item
	If $n^\beta_k < p(p-1)/2$,
	perform a single edge update in $G$ with as proposal to add or remove a uniformly sampled edge with equal probability, if possible:
	\begin{enumerate}
	    \item
	    If $|E| = n^\beta_k$, pick an edge uniformly at random to add to $G$ to obtain the proposed $\tilde{G}$.
	    Set $G = \tilde{G}$ with probability
        \[
            \add{\min\left\{1,\, \frac{\{p(p-1) - 2|E|\}\, p(Y\mid \tilde{G})}{4\, p(Y\mid G)}\right\}}
        \]
	    \item
	    If $|E| = p(p-1)/2$, pick an edge uniformly at random from the $|E|-n^\beta$ free edges in $G$ to remove to obtain the proposed $\tilde{G}$.
	    Set $G = \tilde{G}$ with probability
        \[
            \add{\min \left\{ 1,\, \frac{4\, p(Y\mid \tilde{G})}{(|E| - n^\beta)\, p(Y\mid G)}\right\}}
        \]
	    \item
	    Otherwise,
	    with probability $0.5$,
	    pick an edge uniformly at random from the $p(p-1)/2-|E|$
	    potential edges to add to $G$ to obtain the proposed $\tilde{G}$.
	    Set $G = \tilde{G}$ with probability $\add{\min \{ 1,\, r \, p(Y\mid \tilde{G}) / p(Y\mid G) \}}$
	    where
	    \[
	    r = \begin{cases}
	        \frac{p(p-1) - 2|E|}{|\tilde{E}| - n^\beta}, \quad &|\tilde{E}| = \frac{p(p-1)}{2} \\
	        \frac{p(p-1) - 2|E|}{2(|\tilde{E}| - n^\beta)}, \quad &\text{otherwise}
	    \end{cases}
	    \]
	    With probability $0.5$,
	    pick an edge uniformly at random from the $|E|-n^\beta$ free
	    edges to remove from $G$ to obtain the proposed $\tilde{G}$.
	    Set $G = \tilde{G}$ with probability $\add{\min \{ 1,\, r\, p(Y\mid \tilde{G}) / p(Y\mid G) \}}$
	    where
	    \[
	    r = \begin{cases}
	        \frac{|E| - n^\beta}{p(p-1) - 2|\tilde{E}|}, \quad &|\tilde{E}| = n^\beta \\
	        \frac{2(|E| - n^\beta)}{p(p-1) - 2|\tilde{E}|}, \quad &\text{otherwise}
	    \end{cases}
	    \]
	\end{enumerate}
	\item \label{step:joint}
	For $i=1,\dots,p$:
	\begin{enumerate}
	    \item
	    Sample a proposal $(\tilde{z}_i, \tilde{G})$ from
	    $q(\tilde{z}_i, \tilde{G}\mid z_i, G)$ defined in the text.
	    \item
	    Set $(z_i, G) = (\tilde{z}_i, \tilde{G})$ with probability $\add{\min \{1, R \}}$ where $R$ is given by \eqref{eq:R}.
	\end{enumerate}
	\item
	Update $\nu$ as in Step~\ref{step:nu} of Algorithm~\ref{alg:sbm}.
\end{enumerate}
\end{algorithm}

The joint update for $z$ and $G$ is a Metropolis-within-Gibbs step
where we update $(z_i,G)$ conditional on $z_{-i}$ for $i\in V$ sequentially.
The remainder of this \add{subsection} considers this conditional update where we do not always make the conditioning on $z_{-i}$ explicit.
The target distribution is proportional to
$\pi(z_i, G) = p(z_i\mid z_{-i})\, p(G\mid z)\, p(Y\mid G)
= p(z_i)\, {p(G\mid z_i)}\, {p(Y\mid G)}$
where $p(z_i\mid z_{-i})=p(z_i)$
is given by \eqref{eq:dir}.

We choose a Metropolis-Hastings proposal
that factorizes as $q(\tilde{z}_i, \tilde{G}\mid z_i, G) = {q(\tilde{z}_i\mid z_i)}\, {q(\tilde{G}\mid \tilde{z}_i, G)}$ where $(\tilde{z}_i, \tilde{G})$ denotes the proposed value.
We set ${q(\tilde{z}_i\mid z_i)}$ \add{equal to a uniform distribution over the $(K^{-i} + 1)$ possible values for $\tilde{z}_i$ such that ${q(\tilde{z}_i\mid z_i)} = 1/(K^{-i} + 1)$.
The terms ${q(\tilde{z}_i\mid z_i)}$  and ${q(z_i\mid \tilde{z}_i)}$  cancel out in the acceptance probability, since $z_i$ and $\tilde{z}_i$ lead to the same value of 
 $K^{-i}$}.
We define $q(\tilde{G}\mid \tilde{z}_i, G)$ by
\begin{enumerate}
\item
adding all edges $\{(i,j)\mid z_j = \tilde{z}_i\}$, if any, following from $\tilde{z}_i = z_j = \tilde{k}$ implying $(i,j)\in \tilde{G}$,
and,
\item
if $\tilde{z}_i\ne z_i$,
resampling all edges $\{(i,j)\mid z_j = z_i\}$ with probability $\rho$,
\item
if $\tilde{z}_i = z_i$,
resampling all edges $\{(i,j)\mid z_j \ne z_i\}$ with probability $\rho$.
\end{enumerate}
Since this construction follows the prior $p(G\mid z) = p(G\mid z_i)$,
we have
${q(\tilde{G}\mid \tilde{z}_i, G)} / {q( G\mid z_i, \tilde{G})} = {p(\tilde{G}\mid\tilde{z}_i)} / {p(G\mid z_i)}$
which simplifies the Metropolis-Hastings acceptance probability.
The acceptance probability equals
$\min\{1, R\}$ where
\begin{equation} \label{eq:R}
	R = \frac{q(z_i, G\mid \tilde{z}_i, \tilde{G})\, \pi(\tilde{z}_i, \tilde{G})}{q(\tilde{z}_i, \tilde{G}\mid z_i, G)\, \pi(z_i, G)}
	= \frac{\add{p(\tilde{z}_i)\,} p(Y\mid \tilde{G})}{\add{p(z_i)\,} p(Y\mid G)}
\end{equation}

\subsection{Multiple Graphs}
\label{sec:multi_mcmc}

Algorithm~\ref{alg:multi_sbm} details the MCMC for the proposed model.
For notational convenience, we define $g_{xi}$ also for $x=1$ by $g_{1i}=1$.
Conditionally on $g_{xi}$, it largely mimics Algorithm~\ref{alg:sbm}.
It uses a modified version of \eqref{eq:dir}:
\begin{equation} \label{eq:multi_dir}
	\mathrm{Pr}(z_{1i}=k\mid z \setminus \{z_{xi}\mid g_{xi} = 1 \}) =
	\begin{cases}
		\frac{n_{0k}^{-i}}{\alpha + \sum_{k=1}^{K^{-i0}} n_{0k}^{-i}}, &k=1,\dots,K^{-i0} \\
		\frac{\alpha}{\alpha + \sum_{k=1}^{K^{-i0}} n_{0k}^{-i}}, &k=K^{-i0}+1
	\end{cases}
\end{equation}
where
$n_{0k}^{-i}
= \sum_{j\ne i}\mathds{1}[z_{1j}=k] + \sum_{x\geq 2} \sum_{\{j\mid g_{xj} = 0\}}\mathds{1}[z_{xj}=k]$
and $K^{-i0}$ is the number of unique elements in $z \setminus \{z_{xi}\mid g_{xi} = 1 \}$.
Similarly,
we have for $(i,x)$ such that $g_{xi}=0$,
\begin{equation} \label{eq:multi_dir_x}
	\mathrm{Pr}(z_{xi}=k\mid z \setminus z_{xi}) =
	\begin{cases}
		\frac{n_{k}^{-ix}}{\alpha + \sum_{k=1}^{K^{-i0}} n_{k}^{-ix}}, &k=1,\dots,K^{-ix} \\
		\frac{\alpha}{\alpha + \sum_{k=1}^{K^{-i0}} n_{0k}^{-i}}, &k=K^{-i0}+1
	\end{cases}
\end{equation}
where
$n_{k}^{-ix}
= \sum_j\mathds{1}[z_{1j}=k] + \sum_{\xi\geq 2} \sum_{\{j\mid g_{\xi j} = 0\}}\mathds{1}[z_{j\xi}=k] - \mathds{1}[z_{xi}=k]$
and $K^{-ix}$ is the number of unique elements in $z \setminus z_{xi} =\{z_{1j}\mid 1\leq j\leq p\} \cup \{z_{\xi j}\mid 1\leq j\leq p, 2\leq \xi\leq q\} \setminus z_{xi}$.

\begin{breakablealgorithm}
\caption{MCMC step for the multiple graph stochastic blockmodel. \label{alg:multi_sbm}}
\begin{enumerate}
	\item
	\label{step:G_x}
	For $x=1,\dots,q$, update $G_x\mid \mu,Y$ using WWA from \citet{vandenBoom2022}.
	\item
	For $x=1,\dots,q$, for all $1\leq i<j\leq p$:
	\begin{enumerate}
		\item
		If $(i,j)\in E_x$,
		sample $\zeta_{xij}\sim \mathcal{N}_{(0,\infty)}(\mu_{xij}, 1)$.
		\item
		If $(i,j)\notin E_x$,
		sample $\zeta_{xij}\sim \mathcal{N}_{(-\infty, 0)}(\mu_{xij}, 1)$.
	\end{enumerate}
	\item
	For $k=1,\dots,K$,
	let $S_{kx} = \{(i,j)\mid i < j,\, z_{xi} = z_{xj} = k\}$ ($x=1,\dots,q$) and sample
	$\beta^\star_k \sim \mathcal{N}\{\sigma^2_k \sum_x \sum_{(i,j)\in S_{kx}} (\zeta_{xij} - \theta_i - \theta_j),\, \sigma^2_k\}$
	where $\sigma^2_k = (s_\beta^{-2} + \sum_x |S_{kx}|)^{-1}$.
	\item
	Let $\zeta_{xij}^{-\beta} = \zeta_{xij} - \mathds{1}
[z_{xi}=z_{xj}]\beta_{xi}$ for $1\leq i<j\leq p$ and $x=1,\dots,q$,
and set $\zeta_{xji}^{-\beta} = \zeta_{xij}^{-\beta}$ for notational convenience.
	For $m=1,\dots,M$,
	let $S^\theta_m = \{(i,j)\mid i < j,\, c_i = c_j = m\}$ and sample
	$\theta^\star_m\sim \mathcal{N}[\sigma^2_m \{2\sum_{x,S^\theta_m} \zeta_{xij}^{-\beta} + \sum_{x,\{(i,j)\mid c_i = m \ne c_j\}} (\zeta_{xij}^{-\beta} - \theta_j)\}, \sigma^2_m]$
	where $\sigma^2_m = \{s_\theta^{-2} + 4q|S^\theta_m| + qn^\theta_m(p-n^\theta_m)\}^{-1}$
	with $n^\theta_m = |\{1\leq j \leq p \mid c_j = k\}|$.
	\item
	Update the block labels $z_1$ conditional on the $g_{xi}$. For $i=1,\dots,p$:
	\begin{enumerate}
	\item
	Set $K^{-i0}$ equal to the number of unique values in $z \setminus \{z_{xi}\mid g_{xi} = 1 \}$
	and
	relabel these values such that they are less than or equal to $K^{-i0}$.
	\item
	Denote the $\mu$ resulting from $z_{1i}=k$ by $\mu^k$.
	Here, $\mu^{K^{-i0}+1}$ does not involve the not yet specified $\beta^\star_{K^{-i0}+1}$.
	The neighbourhood of node $i$ in $G_x$ is $N_{xi} = \{j\mid (i,j)\in E_x\}$.
	Sample $z_{1i}$ according to
	\begin{multline*}
		p(z_{1i}=k\mid \text{---}) \\
		\propto
		\mathrm{Pr}(z_{1i}=k\mid \{z_{xi}\mid g_{xi} = 1 \})
		\prod_{\{x\mid g_{xi} = 1\}} \left[\prod_{j\in N_{xi}} \Phi(\mu^k_{xij}) \prod_{j\notin N_{xi}} \{ 1 - \Phi(\mu^k_{xij}) \}\right]
	\end{multline*}
	for $k=1,\dots,K^{-i0}+1$
	where
	$\mathrm{Pr}(z_{1i}=k\mid \{z_{xi}\mid g_{xi} = 1 \})$
	is given by \eqref{eq:multi_dir}.
	\item
	If $z_{1i} \leq K^{-i0}$, set $\beta_{1i} = \beta^\star_{z_{1i}}$ and $K=K^{-i0}$.
	\item
	If $z_{1i} = K^{-i0} + 1$,
	sample $\beta_{1i}\sim\mathcal{N}(0,s^2_\beta)$
	and set $K = K^{-i0} + 1$.
	\item
	Update $z_{xi}$ and $\beta_{xi}$ for $x\geq 2$ such that $g_{xi}=1$ accordingly.
	\end{enumerate}
	\item
	Update the block labels $c$. For $i=1,\dots,p$:
	\begin{enumerate}
	\item
	Set $M^{-i}$ equal to the number of unique values in $c_{-i}$
	and
	relabel $c_{-i}$ such that $1\leq c_j\leq M^{-i}$ for $j\ne i$.
	\item
	Sample $c_i$ according to
	\begin{multline*}
	p(c_i=m\mid \text{---})\propto
	\mathrm{Pr}(c_i=k\mid c_{-i}) \\
	\times
	\begin{cases}
		\exp\left\{
		    \theta^\star_m \sum_{x,j\ne i} (\zeta_{xij}^{-\beta} - \theta_j)
		    - \frac{q(p - 1)}{2}(\theta^\star_m)^2
		\right\}, \quad &k=1,\dots,K^{-i} \\
		\frac{\sigma_c}{s_\theta} \exp\left( \frac{\mu_c^2}{2\sigma_c^2} \right),
		\quad &k = K^{-i} + 1
	\end{cases}
	\end{multline*}
	where
	$\mathrm{Pr}(z_i=k\mid z_{-i})$
	is given by \eqref{eq:dir_theta},
	$\sigma_c^2 = \{q(p-1) + s_\theta^{-2}\}^{-1}$
	and
	$\mu_c = \sigma_c^2\sum_{x,j\ne i}(\zeta_{xij}^{-\beta} - \theta_j)$.
	\item
	If $c_i \leq M^{-i}$, set $\theta_i = \theta^\star_{c_i}$ and $M=M^{-i}$.
	\item
	If $c_i = M^{-i} + 1$,
	sample $\theta_i\sim\mathcal{N}(\mu_c, \sigma_c^2)$
	and set $M = M^{-i} + 1$.
	\end{enumerate}
	\item
	Sample $(g_{xi}, z_{xi})$ from its full conditional: for $i=1,\dots,p$, for $x=2,\dots,q$,
	run Algorithm~\ref{alg:gix}.
	\item
	Let $p' = p + \sum_{x= 2}^q \sum_{i=1}^p \mathds{1}[g_{xi} = 0]$.
	Sample $t_\nu\sim \mathrm{Beta}(\nu + 1,\, p')$.
	Then, sample $\nu$ from the mixture
	$\pi_\nu \Gamma(a_\nu + K,\, b_\nu - \log t_\nu) + (1-\pi_\nu) \Gamma(a_\nu + K - 1,\, b_\nu - \log t_\nu)$
	where $\pi_\nu$ is defined by
	\[
	    \frac{\pi_\nu}{1 - \pi_\nu} = \frac{a_\nu + K - 1}{p'\,(b_\nu - \log t_\nu)}
	\]
	\item
	Update $\alpha$ as in Step~\ref{step:alpha} of Algorithm~\ref{alg:sbm}.
\end{enumerate}
\end{breakablealgorithm}

To update $(g_{xi}, z_{xi})$, $x\geq 2$, according to its full conditional,
we first sample $g_{xi}$
and then $z_{xi}\mid g_{xi}$.
We sample $g_{xi}$ from its conditional distribution
where we condition on everything but $(g_{xi}, z_{xi})$.
To derive this distribution,
consider
\[
\begin{aligned}
    \frac{p(g_{xi} = 1\mid \text{---})}{p(g_{xi}=0\mid \text{---})}
    &=
    \frac{p(g_{xi}=1\mid \beta\setminus\beta_{xi}, g\setminus g_{xi}, G_x)}{p(g_{xi}=0\mid \beta\setminus\beta_{xi}, g\setminus g_{xi}, G_x)} \\
    &= \frac{p(g_{xi}=1\mid \beta\setminus\beta_{xi}, g\setminus g_{xi})
    \, p(G_x \mid \beta\setminus\beta_{xi}, g,g_{xi}=1)}{p(g_{xi}=0\mid \beta\setminus\beta_{xi}, g\setminus g_{xi})
    \, p(G_x \mid \beta\setminus\beta_{xi}, g,g_{xi}=0)}
\end{aligned}
\]
where the last equality follows from Bayes' rule.
Here,
$p(g_{xi}=1\mid \beta\setminus\beta_{xi}, g\setminus g_{xi}) = \gamma$ and $p(g_{xi}=0\mid \beta\setminus\beta_{xi}, g\setminus g_{xi}) = 1 - \gamma$,
and
\[
\begin{aligned}
    p(G_x \mid \beta\setminus\beta_{xi}, g,g_{xi})
    &=
    p(G_x \mid \beta\setminus\beta_{xi}, g) \\
    &= \sum_{z_{xi}} p(z_{xi}, G_x \mid \beta\setminus\beta_{xi}, g) \\
    &= \sum_{z_{xi}} p(z_{xi} \mid \beta\setminus\beta_{xi}, g)\, p(G_x \mid z_{xi}, \beta\setminus\beta_{xi}, g).
\end{aligned}
\]
If $g_{xi} = 0$, $p(z_{xi} \mid \beta\setminus\beta_{xi}, g)$
is equal to \eqref{eq:multi_dir_x}.
If $g_{xi}= 1$,
then $p(z_{xi} \mid \beta\setminus\beta_{xi}, g) = \mathds{1}[z_{xi} = z_{1i}]$.
Finally,
denote the $\mu$ resulting from $z_{xi}=k$ by $\mu^k$.
	Here, $\mu^{K^{-ix}+1}$ does not involve the not yet specified $\beta^\star_{K^{-ix}+1}$.
	The neighbourhood of node $i$ in $G_x$ is $N_{xi} = \{j\mid (i,j)\in E_x\}$.
Then,
\[
    p(G_x \mid z_{xi} = k, \beta\setminus\beta_{xi}, g)
    \propto
    \prod_{j\in N_{xi}} \Phi(\mu^k_{xij}) \prod_{j\notin N_{xi}} \{ 1 - \Phi(\mu^k_{xij})\}
\]
where the normalisation constant does not depend on $g_{xi}$ or $z_{xi}$.
Algorithm~\ref{alg:gix} details the resulting update for $(g_{xi}, z_{xi})$.

\begin{algorithm}
\caption{Full conditional update for $(g_{xi}, z_{xi})$ in Algorithm~\ref{alg:multi_sbm}. \label{alg:gix}}
\begin{enumerate}
	\item
	Flip the value of $g_{xi}$
	Sample $g_{xi}$ according to the distribution
	with as odds of $g_{xi}=1$ versus $g_{xi}=0$ the ratio
	\[
	    \frac{\gamma\, \prod_{j\in N_{xi}} \Phi(\mu^{z_{0i}}_{xij}) \prod_{j\notin N_{xi}} \{ 1 - \Phi(\mu^{z_{0i}}_{xij})\}}{(1-\gamma)\, \sum_{k=1}^{K^{-ix} + 1} \mathrm{Pr}(z_{xi}=k\mid z \setminus z_{xi})\, \prod_{j\in N_{xi}} \Phi(\mu^k_{xij}) \prod_{j\notin N_{xi}} \{ 1 - \Phi(\mu^k_{xij})\}}
	\]
	where $\mathrm{Pr}(z_{xi}=k\mid z \setminus z_{xi})$
	is given by \eqref{eq:multi_dir_x}.
	\item
	\begin{enumerate}
	    \item
	    If $g_{xi} = 0$,
	    update $z_{xi}$:
	    \begin{enumerate}
	\item \label{step:6a}
	Set $K^{-ix}$ equal to the number of unique values in $z \setminus z_{xi}$
	and
	relabel these values such that they are less than or equal to $K^{-ix}$.
	\item \label{step:6b}
	Denote the $\mu$ resulting from $z_{xi}=k$ by $\mu^k$.
	Here, $\mu^{K^{-ix}+1}$ does not involve the not yet specified $\beta^\star_{K^{-ix}+1}$.
	The neighbourhood of node $i$ in $G_x$ is $N_{xi} = \{j\mid (i,j)\in E_x\}$.
	Sample $z_{xi}$ according to
	\[
		p(z_{xi}=k\mid \text{---})\propto
		\mathrm{Pr}(z_{xi}=k\mid z \setminus z_{xi})
		\prod_{j\in N_{xi}} \Phi(\mu^k_{xij}) \prod_{j\notin N_{xi}} \{ 1 - \Phi(\mu^k_{xij}) \}
	\]
	for $k=1,\dots,K^{-ix}+1$
	where
	$\mathrm{Pr}(z_{xi}=k\mid z \setminus z_{xi})$
	is given by \eqref{eq:multi_dir_x}.
	\item
	If $z_{xi} \leq K^{-ix}$, set $\beta_{xi} = \beta^\star_{z_{xi}}$ and $K=K^{-ix}$.
	\item \label{step:6d}
	If $z_{xi} = K^{-ix} + 1$,
	sample $\beta_{xi}\sim\mathcal{N}(0,s^2_\beta)$
	and set $K = K^{-ix} + 1$.
	\end{enumerate}
	    \item
	    If $g_{xi} = 1$,
	    set $z_{xi} = z_{0x}$.
	\end{enumerate}
\end{enumerate}
\end{algorithm}

\section{Model from \texorpdfstring{\citet{Sun2014}}{Sun et al. (2014)}}
\label{sec:sun}

This section describes the model from \citet{Sun2014}
and an MCMC algorithm for it as used in Section~\ref{sec:recovery}.
The prior on the block structure $z$ is the same Chinese restaurant process as in Section~\ref{sec:sics}.
Then,
let
$U=Y^\top Y$
and
define the $p\times p$ matrix $D(z)$
with a block diagonal structure given by $z$
by
$D_{ij}(z) = \mathds{1}[z_i = z_j] W_{ij}/\delta' = \mathds{1}[z_i = z_j] U^{-1}_{ij}$
where the last equality follows from
$W$ being the empirical precision matrix $n U^{-1}$
and $\delta' = \max(p, n) = n$ since $n \geq p$ in Section~\ref{sec:recovery}.
Now,
$\Omega\mid z \sim \mathrm{Wishart}\{D(z), \delta'\}$,
a Wishart distribution with scale matrix $D(z)$ and degrees of freedom $\delta'$.
Finally,
conditionally on $\Omega$,
the rows of $Y$ are independently distributed according to
$\mathcal{N}(0_{p\times 1},\, \Omega^{-1})$.

\citet{Sun2014} use $a_\nu=b_\nu =1$
for the $\Gamma(a_\nu, b_\nu)$ prior on the precision parameter $\nu$ of the Chinese restaurant process.
We set $a_\nu=b_\nu =2$ for consistency with the other methods considered.

\begin{algorithm}[tbp]
\caption{MCMC step for the model from \citet{Sun2014}. \label{alg:sun}}
\begin{enumerate}
	\item
	Update the block labels $z$. For $i=1,\dots,p$:
	\begin{enumerate}
	\item
	Set $K^{-i}$ equal to the number of unique values in $z_{-i}$
	and
	relabel $z_{-i}$ such that $1\leq z_j\leq K^{-i}$ for $j\ne i$.
	\item
	Sample $z_i$ according to
	\[
		p(z_i=k\mid \text{---})\propto
		\mathrm{Pr}(z_i=k\mid z_{-i})\,
		\frac{|D(z)|^{n/2}}{|I_p + D(z)\, U|^{(n+\delta')/2}}
	\]
	for $k=1,\dots,K^{-i}+1$
	where
	$\mathrm{Pr}(z_i=k\mid z_{-i})$
	is given by \eqref{eq:dir}.
	\end{enumerate}
	\item
	Update $\nu$ as in Step~\ref{step:nu} of Algorithm~\ref{alg:sbm}.
\end{enumerate}
\end{algorithm}

Algorithm~\ref{alg:sun} describes an MCMC algorithm targeting the resulting posterior on $z$ based on Algorithm~1 of \citet{Sun2014}.
It does not contain split-merge Metropolis-Hastings updates of $z$ \citep{Jain2004}, unlike Algorithm~1 of \citet{Sun2014},
in line with the other MCMC algorithms presented.
Such split-merge updates can improve the traversal of the MCMC chain between cluster allocations that are separated by configurations with low posterior probability when viewed in terms of updating each $z_i$ in isolation.
These split-merge steps can be added to the MCMC algorithms presented in this supplement though with some added complexity \citep{Dahl2005} as $p(Y\mid z)$ is not available in closed form in the proposed models.
We do not consider this extra complication here as the scenarios of interest to us have sufficient posterior uncertainty in $z$ such that the presented simpler MCMC algorithms are able to effectively explore the posterior on $z$.

\section{Proposed Bayes Factor Computation vs \texorpdfstring{\\}{} Harmonic Mean Approach}
\label{sec:karate_BF}

A motivation for the Bayes factor computation based on the Savage-Dickey ratio from Section~\ref{sec:test}
is that it is converges faster and is more stable than the harmonic mean approach \add{\citep{Newton1994,Raftery2007}}.
Here, we empirically consider that benefit for the simulation study in Section~\ref{sec:karate}.

To assess the differences in convergence and stability,
we inspect how the estimates from both approaches vary
as we increase the number of recorded MCMC iterations these estimates are based on.
The variable part for the proposed approach is the MCMC estimate of $p(z^\star\mid Y)$.
Here,
$z^\star$ corresponds to $K^\star = 1$ block as in Section~\ref{sec:karate}.
The harmonic mean approach
requires the estimation of both
$p(Y\mid \mathcal{M}^\star)$
and
$p(Y\mid \mathcal{M})$
to compute the Bayes factor as defined in \eqref{eq:BF}.
The harmonic mean estimate for $p(Y\mid \mathcal{M}^\star) = p(Y\mid z^\star)$
would involve running a separate MCMC chain with $z$ fixed
to $z^\star$ since $p(Y\mid z^\star)$ is not available in closed form in our model.
Since the latter is not the case if $p(Y\mid z^\star)$ were directly available,
we constrain ourselves to the variability arising from
the estimate for $p(Y\mid \mathcal{M})$ in this comparison.
Since $p(Y\mid z)$ is not directly available,
we take the harmonic mean with the MCMC samples of the Gaussian density $p(Y\mid \Omega)$
as the estimate of $p(Y\mid \mathcal{M}) = p(Y)$.
Note that the precision matrix $\Omega$ is sampled as part of Step~\ref{step:G} in Algorithm~\ref{alg:sbm}.

\begin{figure}[tbp]
\centering
\includegraphics[width=\textwidth]{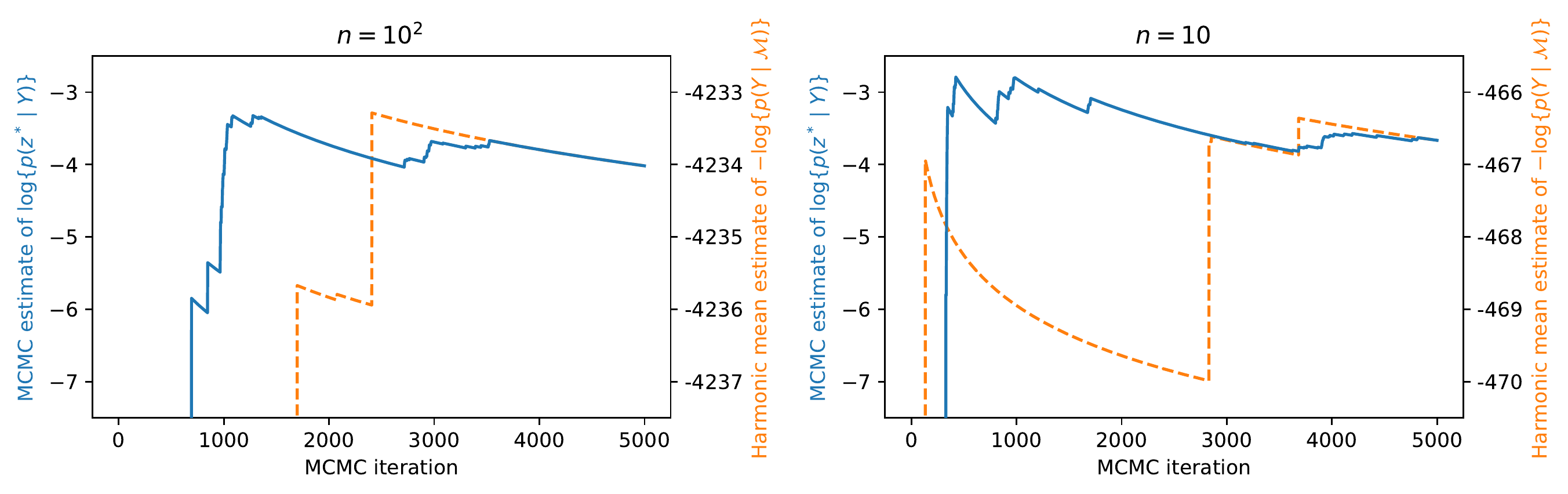}
\caption{
Log of the MCMC estimates constituting the proposed Bayes factor computation (solid lines) and the harmonic mean approach (dashed lines) versus the number of MCMC iterations used for the simulation study in Section~\ref{sec:karate}.
}
\label{fig:karate_BF}
\end{figure}

We only consider the scenarios $n=10^2,10$
as $n=10^4,10^3$
invariably yield $B=0$ with the proposed Bayes factor computation for any number of MCMC iterations used.
The Bayes factor $B$
is proportional to $p(Y\mid z^\star)$
and inversely proportional to $p(Y\mid \mathcal{M})$.
This fact gives rise to the comparison in Figure~\ref{fig:karate_BF}
which shows that the proposed approach converges to $B$ faster, with respect to the number of MCMC iterations,
than the harmonic mean approach.

\begin{figure}[tbp]
\centering
\includegraphics[width=\textwidth]{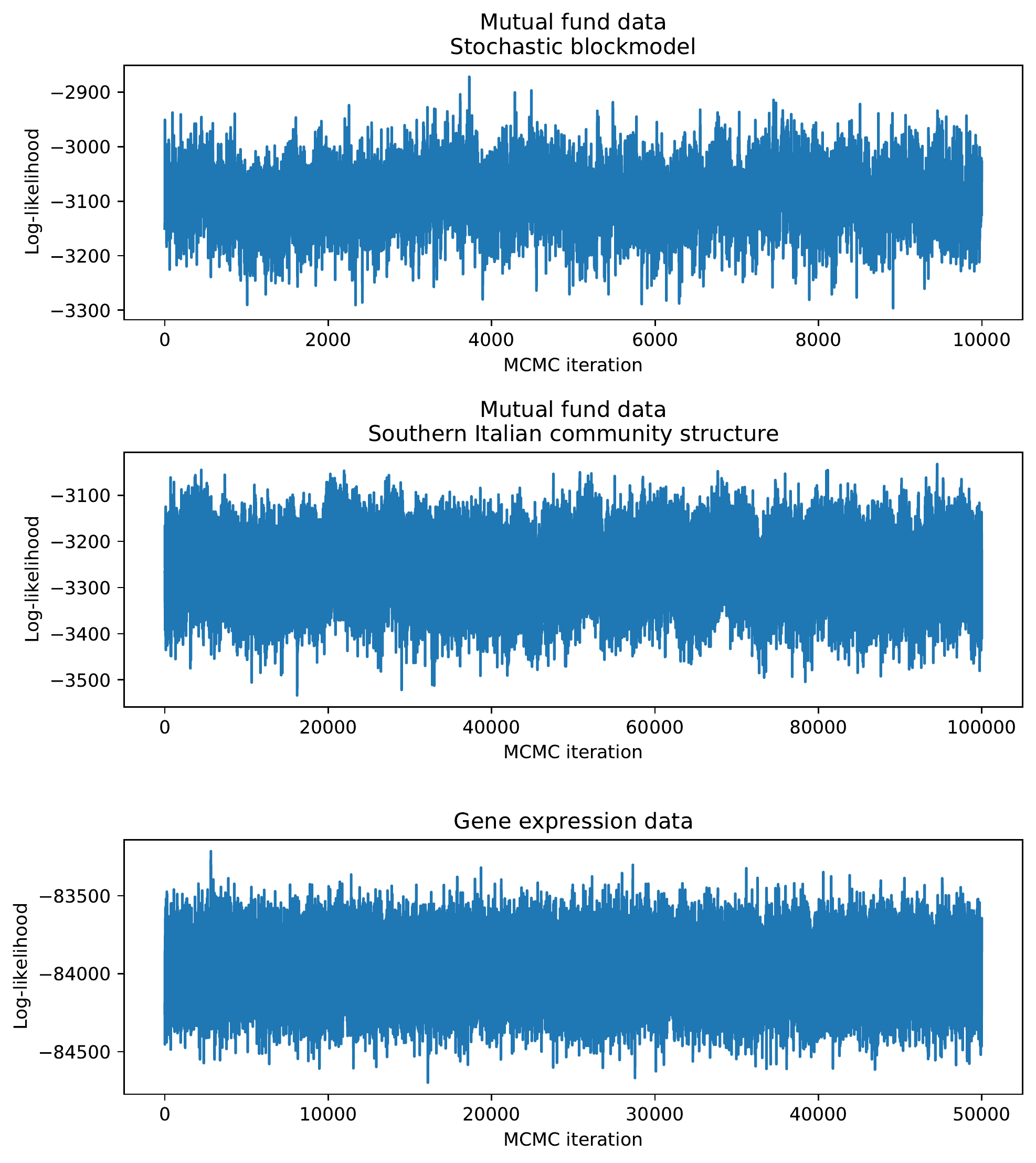}
\caption{
Trace plots of the log-likelihood for the recorded MCMC iterations of the applications in Section~\ref{sec:applications}.
}
\label{fig:trace_plots}
\end{figure}

\section{MCMC Trace Plots}
\label{sec:trace_plots}

To assess the MCMC convergence and mixing for the applications in Section~\ref{sec:applications},
we inspect trace plots.
Specifically,
we consider trace plots of the Gaussian likelihood
$p(Y\mid \Omega)$
in the application with the mutual fund data set.
Recall that the precision matrix $\Omega$ is sampled as part of Step~\ref{step:G} in Algorithm~\ref{alg:sbm}.
For Algorithm~\ref{alg:sics}, we add a step
that samples $\Omega\mid G,Y\sim \mathcal{W}_G(\delta^\star, D^\star)$
at each iteration.
For the multiple graph application with the gene expression data,
we define the precision matrix $\Omega_x$ for each group~$x$
analogously to $\Omega$ in the single graph case.
Then,
the likelihood follows as
$\prod_{x=1}^q p(Y_x\mid \Omega_x)$
where each $p(Y_x\mid \Omega_x)$ is the density resulting from the rows of $Y_x$ being independently distributed according to $\mathcal{N}(0_{p\times 1}, \Omega_x^{-1})$.
Analogously to Algorithm~\ref{alg:sbm},
$\Omega_x$ is sampled as part of Step~\ref{step:G_x} of Algorithm~\ref{alg:multi_sbm}.

Figure~\ref{fig:trace_plots} contains the resulting trace plots.
They do not suggest deficient MCMC convergence or mixing.

\bibliographystyle{chicago}
\bibliography{graph}